\documentclass[sigconf,screen]{acmart}
\AtBeginDocument{%
  }

\usepackage{booktabs}
\usepackage{makecell}
\usepackage{bm}
\usepackage[table]{xcolor}

\setcellgapes{1pt}
\makegapedcells

\setcopyright{acmlicensed}
\copyrightyear{2026}
\acmYear{2026}
\acmDOI{XXXXXXX.XXXXXXX}
\acmConference[CHI '27]{ACM Conference on Human Factors in Computing Systems}{May 10--14,
  2027}{Pittsburgh, PA}
\acmISBN{978-1-4503-XXXX-X/2018/06}

\begin{document}

\title{``I don't know what I don't know'': Exploring the Benefits and Challenges of AI in Interdisciplinary Research}

\author{Indu Panigrahi}
\correspondingauthor
\email{indup2@illinois.edu}
\orcid{0009-0002-2084-4756}
\affiliation{%
  \institution{Siebel School for Computing and Data Science\\University of Illinois Urbana-Champaign}
  \city{Champaign}
  \state{Illinois}
  \country{USA}
}

\author{Tal August}
\email{taugust@illinois.edu}
\orcid{0000-0001-6726-4009}
\affiliation{%
  \institution{Siebel School for Computing and Data Science\\University of Illinois Urbana-Champaign}
  \city{Champaign}
  \state{Illinois}
  \country{USA}}

\renewcommand{\shortauthors}{Panigrahi et al.}

\begin{abstract}
  Interdisciplinary research is widely lauded as beneficial for society but presents unique challenges to the researchers who bridge disciplinary boundaries. Despite the rapid increase in the availability and usage of AI systems for research tasks, there is little work studying how AI systems have impacted workflows that span unfamiliar disciplines. Through a survey (N=$62$) and follow-up interviews (N=$23$) with interdisciplinary researchers, we describe the unique challenges and opportunities for AI in interdisciplinary research workflows. Participants viewed AI both as a solution and exacerbation of the general challenge of ``unknown unknowns'': AI systems could facilitate a broad awareness of a new research area, yet also provide an illusion of understanding. Crucially, participants lacked the expertise and intuition to catch these illusions. Our findings motivate a focus on designing AI systems to support rather than automate collaboration across disciplines, scaffold learning opportunities, and include avenues for triangulation between resources.
\end{abstract}


\newcommand{\ta}[1]{\textcolor{purple}{}} 




\maketitle

\section{Introduction}


The benefits of interdisciplinary research are all around us.
The device that you may be using to read this paper resulted from decades of researchers communicating across disciplines including Computer Science, Chemistry, Policy, Ergonomics, and Sociology \cite{motivation}.
Crossing disciplinary boundaries has become increasingly common in certain fields \cite{Liu_Rong_2026,Li2023interdisc,Smye_Frangi_2021,Klein2008evalid} and has had far-reaching impacts, such as in Structural Biology \cite{Rangan2024cryodrgn}, History (e.g., digital humanities) \cite{Graziosi2023classics}, and Environmental Science \cite{Karns2026env}. 
The history of HCI is itself an adaptation of methods and findings from distinct disciplines (e.g., Psychology, Sociology, Computer Science) into a new study of how humans interact with and through technology \citep{Grudin1990hcihistory,Olson_Olson_2003_hcihistory}.

While interdisciplinary research is widely lauded as beneficial, it also presents unique challenges \citep{leahey2017prominent}. For example, researchers venturing out of their discipline often encounter discipline-specific terminology \citep{lucy-etal-2023-words, Bao2025Words} and expectations surrounding methodological rigor (e.g., standards for inter-rater reliability \cite{McHugh2012diffkappa}). While there is a general consensus that interdisciplinary research workflows are different from in-discipline workflows \citep{palmer1999interdisciplinary,Klein1991interdisc,Barley2022linguistic, Palmer2001workatbound}, there has been little work mapping these differences to tangible differences in the tools (e.g., literature search, document writing) that interdisciplinary workflows use.

This question is especially urgent given the general upheaval of scholarly tools that AI has precipitated \citep{liao2025llms}. The past few years has seen a rapid increase in the availability and usage of AI for research tasks, spanning literature search \citep{August2023paperplain,Fok2023scim,singh2025scholarqa,Whitfield2023elicit}, data analysis \citep{liao2025llms,torres2026aimeetsscienceresearch}, and writing \citep{Jakesch2023cowriting,Monge2025cowrite,Gruda2024writing}. While work has studied how researchers have adopted AI into their workflows in general \citep{liao2025llms, Morris2023ScientistsPO}, there is no work specifically studying how researchers perceive the way AI systems have impacted, or will impact, workflows that span new, unfamiliar disciplines. This perspective is important given the unique aspects of interdisciplinary workflows.
In this paper, we explore the roles AI systems play in interdisciplinary research by answering the following research questions:
\begin{itemize}
    \item \textbf{RQ1:} What challenges do interdisciplinary researchers face when exploring unfamiliar disciplines?
    \item \textbf{RQ2:} How and why do interdisciplinary researchers use AI systems in their research?
    \item \textbf{RQ3:} What challenges do interdisciplinary researchers face when using AI systems in their research?
\end{itemize}

We conduct a survey of $62$ researchers (spanning $102$ disciplines and subdisciplines) to understand their challenges when conducting interdisciplinary research and identify the role that scholarly tools, including those enabled by AI, play in overcoming these challenges and introducing new ones. 
We followed this survey with 60-minute semi-structured interviews with $23$ respondents, in which we explored the challenges that these researchers face when working across disciplines, their current approaches to these challenges, and the potential benefits and risks of AI-enabled tools.

Our findings paint a consistent picture of challenges unique to interdisciplinary work and opportunities that these challenges present for new AI-enabled tools.
As $5$ interview participants stated, ``\textit{I don't know what I don't know}'' when exploring unfamiliar disciplines.
This challenge of the ``unknown unknown'' \citep{bates1989design, Mcmanus2006AFF} found its way into searching for information, aligning expectations with collaborators, and communicating across different communities (Section \ref{sec:rq1_results}). Participants viewed AI both as a solution (Section \ref{sec:rq2_results}) and escalation (Section \ref{sec:rq3_results}) of this general challenge. AI systems could be useful for facilitating a broad understanding of a new research area and for communicating ideas across disciplines. At the same time, participants raised concerns about AI providing an illusion of understanding and lacking taste when exploring an unfamiliar discipline. Crucially, participants lacked the discipline-specific expertise and intuition \citep{palmer1999interdisciplinary,Ballantyne_2019} to catch these gaps.
At a higher level, participants also worried that lowering the barriers between disciplines (e.g., via AI-facilitated search) could conversely lower outside researchers' respect of a discipline.
Participants described ways that they mitigate these issues, including triangulating information between AI, collaborators, and paper reading. 
Our findings motivate a focus on designing AI systems to support rather than automate collaboration across disciplines, scaffold learning opportunities, and include avenues for triangulation.


\section{Related Work}
\subsection{Interdisciplinary Research Workflows}
\label{sec:rw_workflows}
Different threads of work classify interdisciplinary science differently \citep{Klein1991interdisc, huutoniemi2010analyzing, Vilhena_2014}, but most classifications agree that interdisciplinary research draws on or connects different bodies of knowledge to answer a shared question \citep{Daniel2022ChallengesFI} or gain a broader understanding of a phenomenon \citep{huutoniemi2010analyzing}. While scholars have knowledge they can apply outside their discipline (e.g., how a research paper is structured), knowledge of discipline-specific concepts and methodologies is often not shared between disciplines \citep{Choi2007MultidisciplinarityIA, Wudarczyk2021BringingTR, Leahey_2018, yao2025interdisciplinaryteamsfailsystematic}, presenting unique challenges.

Past work has explored and articulated the challenges in interdisciplinary work, particularly for information seeking tasks. For example, \citet{palmer1999interdisciplinary} interviewed interdisciplinary scientists affiliated with an interdisciplinary institute about how they seek and exchange information. Palmer found that researchers often consulted in-discipline scientists and learned directly from presentations and papers. Similarly, \citet{Agapie2022hcihealth} interviewed HCI faculty working with healthcare and identified several challenges that they faced throughout different stages of projects. 

We categorize these experiences from prior work into three dimensions as a structure for our survey and interviews (Section \ref{sec:survey_methods}): Navigating Information, Assessing Information, and Connecting Across Boundaries (Table \ref{tab:categorized_workflow}).
Navigating Information focuses on finding information and understanding the information (e.g., unfamiliar terminology \cite{Agapie2022hcihealth,guo2024personalized,palmer1999interdisciplinary,Spanner2001bordercrossings,Newby2011unfamiliar,foster2004nonlinear,Lee2015moca,Qin1998types}), while Assessing Information is focused on deciding if and how the information found aligns with what the researcher needs (e.g., understanding when information is valuable and complete \cite{Kim2025autoeth,foster2004nonlinear,Zheng2024disciplink}).
Lastly, Connecting Across Boundaries covers experiences that explicitly involve integrating information and resources from different disciplines, such as communicating with collaborators \cite{Wan2026knit,palmer1999interdisciplinary,Cao2025medai,Agapie2022hcihealth,Palmer2001workatbound,Lee2015moca,Qin1998types} and reconciling different expectations surrounding methodology, analysis, and evidence \cite{Cao2025medai,Wan2026knit,Palmer2001workatbound,Colusso2019sciencemodel,Agapie2022hcihealth,Cibrian2022Interventions,Mitchell2021woz,palmer1999interdisciplinary}. The categories are not completely independent as some experiences can affect experiences in other categories (e.g., differences in terminology can affect communication with collaborators), but they serve as a way to organize components of interdisciplinary workflows and structure our study.





\begin{table*}[h]
\sffamily
\small
  \caption{\textbf{Categorization of experiences in interdisciplinary workflows.}}
  \label{tab:categorized_workflow}
  \Description{A table showing 3 rows that categorize the interdisciplinary workflow and 2 columns for each category and its experiences and relevant literature.}
  \begin{tabular}[b]{ll}
    \textbf{Category} & \textbf{Experiences} \\
    \midrule

    Navigating Information & \makecell[lc]{
    Unfamiliar terminology or jargon \cite{Agapie2022hcihealth,guo2024personalized,palmer1999interdisciplinary,Spanner2001bordercrossings,Newby2011unfamiliar,foster2004nonlinear,Lee2015moca,Qin1998types}\\
    Deciding what to ask or search for \cite{Zheng2024disciplink,foster2004nonlinear,Palmer2017inforesearch,palmer1999interdisciplinary,Klein1991interdisc,Qin1998types,Kim2025autoeth}\\
    } \vspace{0.5mm}\\

    \arrayrulecolor{gray}\hline
    Assessing Information & \makecell[lc]{
    Considering a work’s context in its field \cite{Cao2025medai,guo2024personalized,Qin1998types}\\
    Deciding when you have sufficient information \cite{foster2004nonlinear,Zheng2024disciplink}\\
    Recognizing when information is valuable \cite{Kim2025autoeth,foster2004nonlinear} \\
    } \vspace{0.5mm}\\
    
    \arrayrulecolor{gray}\hline
    Connecting Across Boundaries & \makecell[lc]{
    Communication with collaborators \cite{Wan2026knit,palmer1999interdisciplinary,Cao2025medai,Agapie2022hcihealth,Palmer2001workatbound,Lee2015moca,Qin1998types} \\
    Different methodological expectations \cite{Cao2025medai,Wan2026knit,Palmer2001workatbound,Colusso2019sciencemodel,Agapie2022hcihealth,Cibrian2022Interventions,Mitchell2021woz,Klein2008evalid} \\
    Different expectations surrounding analysis and evidence \cite{palmer1999interdisciplinary,Wan2026knit,Agapie2022hcihealth,Klein2008evalid} \\
    Different writing styles \cite{bhatt2025borderlands,palmer1999interdisciplinary}
    } \vspace{.5mm}\\
    
    \arrayrulecolor{black} \bottomrule
  \end{tabular}
\end{table*}



\subsection{AI for Research}
With the rising capabilities and availability of AI in assisting with different stages of research, many researchers have integrated AI into their workflows \cite{liao2025llms,Kapania2025hcillm,Rust2025llmsummary,torres2026aimeetsscienceresearch}.
For example, many AI-powered interfaces have been developed to help search for papers \cite{mudd2025screen}, skim papers \cite{Fok2023scim}, aggregate information across multiple documents \cite{singh2025scholarqa,Whitfield2023elicit}, and understand content \cite{August2023paperplain,Rust2025llmsummary,Fok2023scim,rust2025lifestyle,fok2023qlarify}.
A popular option has been conversation-based systems (i.e., chatbots) that provide summaries of topics and individual papers, such as Elicit \cite{Whitfield2023elicit}, ASTA \cite{singh2025scholarqa}, and OpenAI's Deep Research \cite{ChatGPTDeepResearch}.

Research on AI tools for these uses has surfaced both benefits and challenges.
Challenges include verifying information provided by AI \cite{Jia2026llmcheck,Gu2024aidataanalysis,Kim2025autoeth,Kambhamettu2025traceable,wadden2022multivers,Fecher2025llmimplications} and AI limiting the amount or type of information shown \cite{Sharma2024Echo,Zheng2024disciplink,Kim2025autoeth,Mittal_Blodgett_Liao_2026}.
Benefits include serendipitously encountering useful information from AI \cite{duck2025seren,Chen2025serenspark,Storey2025genaiinfo} and synthesizing information across multiple sources \cite{kambhamettu2026litpivot,liao2025llms,Kapania2025hcillm}.

Prior work has also explored how researchers integrate AI into their workflows \cite{Kapania2025hcillm,Rust2025llmsummary}.
For example, \citet{liao2025llms} surveyed researchers to understand when researchers use LLMs (e.g., information seeking, ideation) and how their backgrounds affect their usage and perceptions. We take inspiration from these works but shift the focus towards the experiences of interdisciplinary researchers specifically.


\subsection{AI for Interdisciplinary Research}

Several systems have been built that integrate AI to improve interdisciplinary workflows specifically.
For example, some work focuses on technical aspects like helping researchers decide what to ask when seeking information \cite{Zheng2024disciplink}, mitigating jargon barriers \cite{guo2024personalized}, and styling writing \cite{bhatt2025borderlands}. For example, \citet{Bao2025Words} developed a prototype cross-domain search engine that used terminology as a way to surface differences between communities (e.g., using different terms for similar concepts). Other work has introduced AI-mediated systems that facilitate collaboration in interdisciplinary teams, such as KNIT \cite{Wan2026knit} and MedAI-SciTS \cite{Cao2025medai}.

To a lesser degree, some work has explored when and how interdisciplinary researchers adopt AI systems.
\citet{Kim2025autoeth} conducted an autoethnography as researchers in biology, information science, and HCI using an AI chatbot to explore unfamiliar topics.
The authors encounter some of the experiences that we categorize under Navigating Information and Assessing Information (Table \ref{tab:categorized_workflow}), such as unfamiliar terminology and recognizing when information is valuable.
However, because the autoethnography focused on trying a specific chatbot (i.e., Llama 3 \cite{llama3modelcard}) to learn a new topic, mostly in isolation, there remain several unexplored experiences of interdisciplinary workflows, particularly those that we categorize under Connecting Across Boundaries (Table \ref{tab:categorized_workflow}).
To the best of our knowledge, no work formally describes how interdisciplinary researchers use and perceive AI, which is our main contribution in this work, surveying several interdisciplinary researchers.

\section{Procedure}
To get an overview of what challenges researchers face in interdisciplinary research and when they find AI useful, we began by deploying a survey that asked about the experiences we identified from prior work (Section \ref{sec:rw_workflows}).
We then conducted semi-structured interviews with a subset of respondents to gain a richer understanding of why the experiences were challenging, the specific ways in which they use AI, and what they perceive to be the benefits and challenges of using AI. All components were reviewed and deemed exempt by our institution's IRB.

\subsection{Study Materials}

\subsubsection{Survey}
\label{sec:survey_methods}
We conduct a survey ($N=62$) to understand what experiences interdisciplinary researchers have when exploring disciplines beyond their primary discipline and when they use AI.
To probe at interdisciplinary researchers' workflows, we asked about experiences in navigating and assessing information and connecting across disciplines that we identified from prior work (Section \ref{sec:rw_workflows}). The survey questions consisted of three main parts; the individual questions are provided in Appendix \ref{sec:survey_questions}:
\begin{enumerate}
    \item \textbf{Background} questions aimed at gathering an overview of respondents' experience with AI and professional background, such as the discipline(s) the respondent has experience in and general AI usage and perceptions. These questions were adapted from prior surveys on researchers' AI usage \cite{kim2024notsure,liu2026writing}.
    \item \textbf{Interdisciplinary experiences} questions about the frequency and difficulty of the experiences listed in Table \ref{tab:categorized_workflow} along with the possible usefulness of AI for addressing those experiences. As described in Section \ref{sec:rw_workflows}, the experiences were synthesized across prior work that explores workflows of interdisciplinary researchers. We also included optional open-ended questions for respondents to list additional interdisciplinary experiences and describe any benefits and challenges they perceived when using AI.
    \item \textbf{Demographic information} questions such as age and gender; these were optional.
\end{enumerate}
We conducted 3 pilots through which we iterated on the wording of questions and shortened the survey to focus on the interdisciplinary experiences. 
We deployed the final survey on Qualtrics.
Participants took an average of $43.48 \pm 133.97$ minutes to complete the survey and could opt in to a raffle for one of five $\$20$ Amazon gift cards at the end of the survey.

\subsubsection{Follow-up Interviews}
To better contextualize and understand the survey responses, we conducted semi-structured interviews with $23$ respondents who marked their interest in the survey.
Each interview lasted 1 hour and was conducted over Zoom.
$22$ participants were compensated with a $\$15$ Amazon Gift Card each; $1$ participant declined the interview compensation.

We began each interview by asking about the participant's motivation for working with disciplines outside of their primary discipline(s), the tasks that their work typically involves, and the resources and tools that they use.
Displaying the participant's survey response for their reference, we asked participants to describe what, if anything, is challenging about the different interdisciplinary experiences (Table \ref{tab:categorized_workflow}).
Finally, to understand their usage and perceptions of AI, we asked about the types of AI systems that they use (e.g., text summaries, coding agents) and the situations in which they use or avoid AI, along with the reasons for each.
The full interview guide can be found in Appendix \ref{sec:int_guide}.

\subsection{Recruitment and Participants}
\label{sec:participants_recruitment}

We advertised across several platforms and collected survey responses during July and August 2026. 
We reached out to venues where we expected interdisciplinary work to be happening \cite{palmer1999interdisciplinary,Wan2026knit,Barley2022linguistic}: emailing departments and interdisciplinary centers across institutions, posting on social media platforms (BlueSky, LinkedIn), emailing interdisciplinary researchers in the authors' networks, and posting on university listservs and newsletters.
The authors have experience conducting interdisciplinary research and were consequently able to circulate recruitment messages among interdisciplinary connections. We considered researchers who have experience or professional interest engaging with an unfamiliar discipline or a discipline beyond their primary discipline to be interdisciplinary researchers for the purposes of our survey and interview \cite{bhatt2025borderlands,Bao2025Words}. We take this broad approach because we were interested in both researchers who formally collaborate across disciplines (e.g., researchers who hold positions in two departments) \cite{palmer1999interdisciplinary,foster2004nonlinear,Barley2022linguistic}, as well as those who have interest doing so. We also let survey respondents self-describe their disciplines and considered both discipline (e.g., Computer Science, Chemistry) and subdiscipline (e.g., HCI, Computer Vision) crossing as interdisciplinary, following prior work \citep{bhatt2025borderlands} and in recognition of the fact that definitions of ``discipline'' are often fuzzy \citep{Strober2006HabitsOT}.

\begin{figure}
    \centering
    \includegraphics[width=\linewidth]{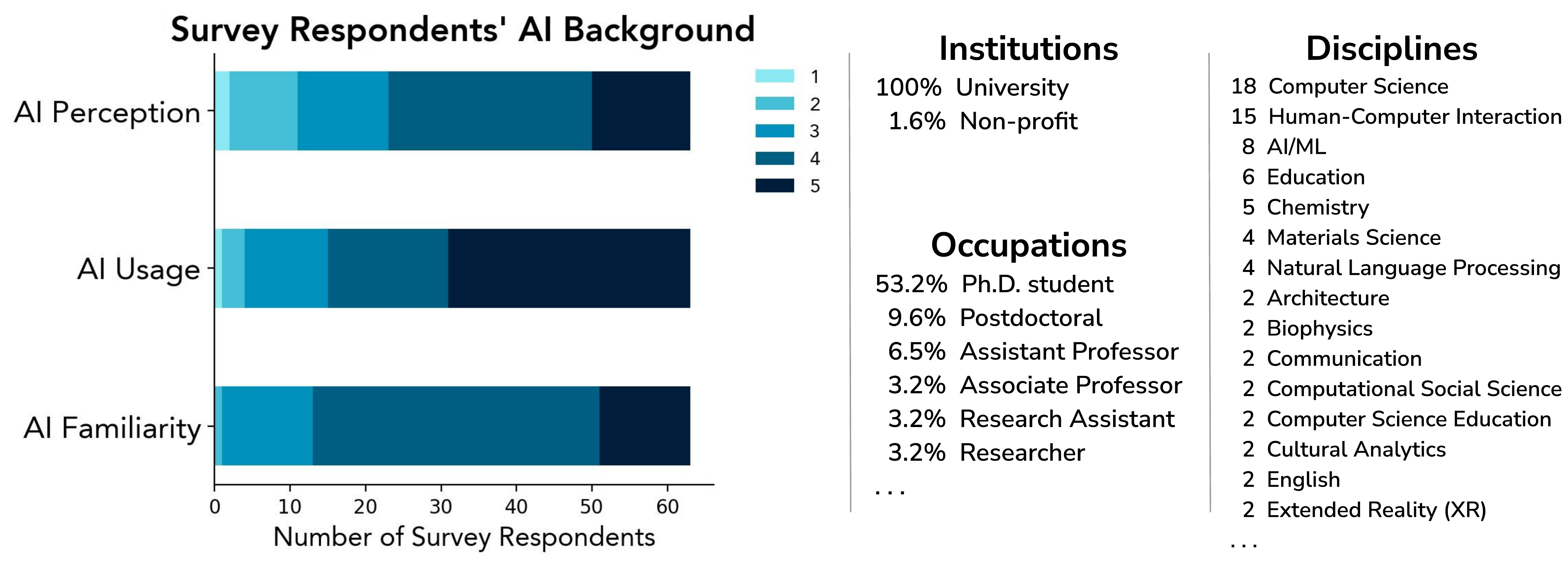}
    \caption{\textbf{An overview of respondents' background}: the left panel shows the distribution of their perceptions of, usage of, and familiarity with AI. Perception ranged from 1 = Negative, 2 = Somewhat negative, 3 = Neutral, 4 = Somewhat positive, and 5 = Positive. Usage ranged from 1 = Never, 2 = Rarely, 3 = Occasionally, 4 = Regularly, and 5 = Frequently. Familiarity ranged from 1 = Not familiar, 2 = Slightly familiar, 3 = Moderately familiar, 4 = Very familiar, and 5 = Extremely familiar. The middle panel lists the distributions of self-reported institutions and occupations that the respondents work in; most were from universities, with a majority being doctoral students. The right panel lists the top fifteen disciplines by mention; participants could list up to five disciplines.}
    \label{fig:respondent_background}
    \Description{On the left, a plot with 3 bars showing the distributions of general familiarity with AI, usage of AI, and perceptions of AI. In the middle, percentages representing the distribution of participants across institutions and occupations. On the right, a list of the top 15 disciplines by number of respondents.}
\end{figure}

Figure \ref{fig:respondent_background} summarizes respondents' experience with AI, institutional affiliations, occupations, and disciplines.
All respondents worked in universities, with the majority being doctoral students ($53.2\%$), influenced by the authors' networks and by interdisciplinary institutes primarily being housed within universities.
Most respondents were involved in STEM disciplines, and $32$ marked having the most years of experience in Computer Science or an area within Computer Science (e.g., HCI) compared to the other disciplines they work with.
In total, respondents covered $102$ disciplines across the humanities, STEM, and social sciences (Appendix \ref{sec:app_participant_background}).
Respondents' demographics are summarized in Appendix \ref{sec:app_participant_background}.
Respondents were generally familiar with AI, most using it regularly ($Mdn=4.0, IQR=4.0-3.0$); general impressions of AI varied from negative to positive ($Mdn=3.0, IQR=3.0-2.0$) (Figure \ref{fig:respondent_background}).

\subsection{Analysis}
\label{sec:analysis}

\subsubsection{Filtering Survey Responses}
We performed post-hoc filtering to ensure we had valid survey responses.
Respondents were asked to link their scholar profile (either Google Scholar, Semantic Scholar, ORCiD, or dblp). We excluded responses that either lacked a profile (1 response) or linked an invalid scholar profile (e.g., fake Google Scholar profiles, links that were not scholar profiles).
After exclusion, we had $62$ valid responses, reduced from $216$ responses received.

\subsubsection{Quantitative Analysis}
We analyze the distributions of ratings for the interdisciplinary experience questions to understand how frequent and challenging different experiences were and how useful AI was perceived to be for each. Figure \ref{fig:survey_results} depicts the distributions, and we report the median and interquartile range when describing specific distributions in the results.

\subsubsection{Qualitative Analysis}
\label{sec:qual_analysis}
To gain a detailed understanding of the challenges that interdisciplinary researchers face and the benefits and challenges that AI presents in those situations, we performed reflexive thematic analysis as outlined by Braun and Clarke \cite{Braun_Clarke_2019,Braun_Clarke_2006_ta} to analyze the qualitative data from the interviews and from the open-ended survey questions.
The interviews were recorded, transcribed, and anonymized for analysis; survey responses were also anonymized.
The author who conducted the interviews familiarized herself with the interviews by listening to each audio recording alongside its transcript and taking preliminary notes while doing so.
The same author then coded the interviews, comparing codes with another author on one interview to check interpretations, and subsequently refined the codes over the course of two weeks.
From the codes, the author constructed and iterated on the themes with the other author for a month before settling on the final set and naming.

Reflecting on code creation and theme construction, the author coding the interviews noted that her previous experience working between Geosciences and Computer Science facilitated her interpretation of the personal experiences that participants described.
More broadly, being researchers in AI and HCI allowed the authors to consider the data in terms of understanding how AI is being integrated into different experiences that the participants described.
The authors also typically conduct mixed-methods research with a systems focus which may have shaped the themes towards design recommendations for AI systems (e.g., ``supporting triangulation between AI output and traditional resources'').

To differentiate between the open-ended survey responses and the interview responses, we append a ``-S'' to the participant ID (e.g., P1-S refers to the survey data from interview participant P1).
For the survey data from survey-only respondents, we use ``S'' for the ID (e.g., S1 refers to a survey-only respondent, not the survey data for P1).

\section{Survey Results}
\label{sec:survey_results}

\begin{figure}[h]
    \centering
    \includegraphics[width=\linewidth]{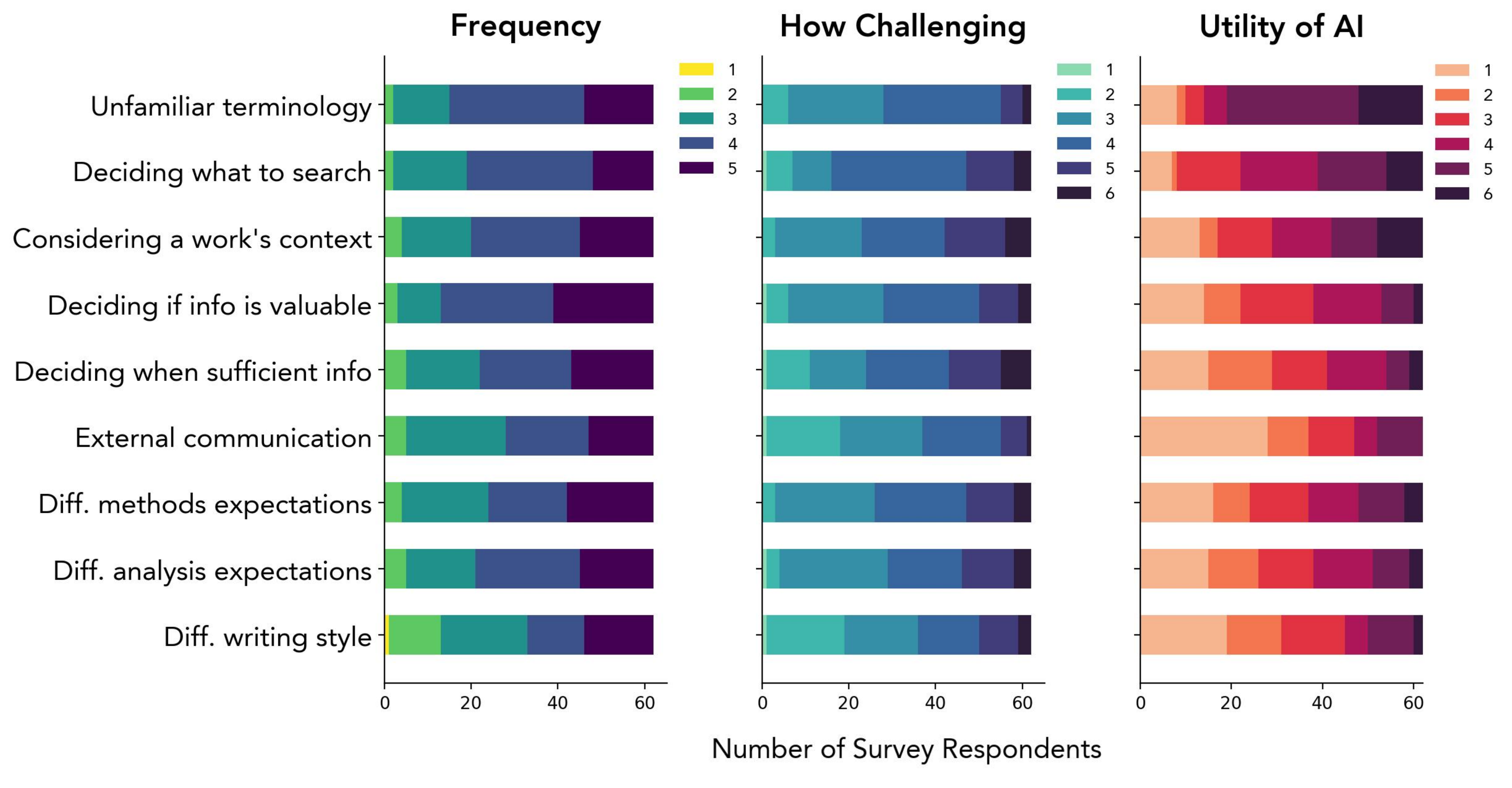}
    \caption{\textbf{The frequency of, difficulty of, and utility of AI for interdisciplinary experiences across the $\mathbf{62}$ survey respondents.} Each row (e.g., ``Unfamiliar terminology'') represents an experience question from our survey (Table \ref{tab:categorized_workflow}). The rating scales were as follows: Frequency (1 = Never, 2 = Rarely, 3 = Occasionally, 4 = Regularly, and 5 = Frequently); How Challenging (1 = I do not experience this, 2 = Not challenging, 3 = Slightly challenging, 4 = Moderately challenging, 5 = Very challenging, 6 = Extremely challenging); Utility of AI (1 = I do not use AI for this, 2 = Not useful, 3 = Slightly useful, 4 = Moderately useful, 5 = Very useful, 6 = Extremely useful). The corresponding questions are provided in Appendix \ref{sec:interdisc_exp_questions}.
    }
    \label{fig:survey_results}
    \Description{3 stacked bar plots showing the distributions of frequency, challenging, and AI usage for different experiences in interdisciplinary research.}
\end{figure}

Figure \ref{fig:survey_results} shows the distributions of survey responses for each interdisciplinary experience (e.g., ``Unfamiliar terminology'') in each question category (e.g., ``Frequency'').

Respondents rated deciding when information is valuable as the most frequent experience ($Mdn = 3.0, IQR = 4.0-3.0$), followed by unfamiliar terminology ($Mdn = 3.0, IQR = 3.75-3.0$).
In contrast, respondents encountered different writing styles less frequently than the other experiences ($Mdn = 2.0, IQR = 3.75-2.0$).

For difficulty, respondents rated considering a work's context ($Mdn = 3.0, IQR = 4.0-2.0$) as more challenging than all experiences except deciding what to search for ($Mdn = 3.0, IQR = 3.0-2.25$).
Different writing styles ($Mdn = 2.0, IQR = 3.0-1.0$) and communication with other disciplines ($Mdn = 2.0, IQR = 3.0-1.0$) were rated as less challenging than the other experiences.

Lastly, participants found AI most useful for addressing unfamiliar terminology ($Mdn = 4.0$, $IQR = 4.0-3.0$), followed by deciding what to search for ($Mdn = 3.0$, $IQR = 4.0-2.0$), suggesting that AI is particularly useful for navigating information (Table \ref{tab:categorized_workflow}).
AI was rated relatively less useful for communication with others ($Mdn = 1.0$, $IQR = 2.0-0.0$) and different writing styles ($Mdn = 1.5$, $IQR = 3.0-0.0$).
That being said, the utility of AI for all experiences exhibited some degree of variance, and we discuss these nuances in our qualitative results (Section \ref{sec:qual_results}).

\section{Qualitative Results}
\label{sec:qual_results}

\subsection{Setting the scene: Participant background and experience with AI}
Generally participants reported having one or more disciplines they had varying degrees of experience in; Table \ref{tab:interview_participant_background} in Appendix \ref{sec:app_participant_background} provides a list of participants and their years of experience in each discipline. Participants referenced many ways in which they viewed their work as interdisciplinary. These included seeking out other disciplines as applications of their primary discipline (e.g., ``\textit{The political science and marketing collaborations are using methods that I've already worked on developing}'' - P17), providing advice to out-of-discipline researchers (e.g., ``\textit{we do a lot of compliance work, trying to assess the potential impact of, say, expanding a highway in an area on cultural resources}'' - P19), and the research questions they were pursuing as being inherently interdisciplinary (e.g., ``\textit{how people teach writing is as intertwined to the research as writing itself}'' - P8).

In terms of traditional non-AI resources, participants described using resources of different media from academic papers (P1-23) to physical items (e.g., ``soil samples'') (quoted from P19, echoed by P11, P15) to datasets (e.g., IMLS database\footnote{https://www.imls.gov/research-evaluation/data}) (from P12, echoed by P10, P19).
A few participants mentioned resources that allow for passive communication and learning, such as observing how members of other disciplines interact on social media (P17, P22), converse or describe concepts on podcasts (P16) and at gatherings (P21, P22), and revise writing (P22). As P22 wrote in their survey response, ``\textit{Social media has been useful for me when entering new fields. I can follow along with `insider' conversations and learn passively over time.}''
When using AI, all participants mentioned querying through chatbot interfaces (e.g., ChatGPT, Claude, Gemini), with a subset using coding agents (e.g., Claude Code, Copilot, Codex) (P5, P6, P8, P12, P18, P20, P22, P23). While most participants used proprietary models, P1 and P19 also mentioned using local models.



\begin{table*}[h]
\sffamily
\small
  \caption{\textbf{Summary of challenges in interdisciplinary research along with the usage, benefits, and challenges of AI for interdisciplinary researchers.}}
  \label{tab:results_summary}
  \Description{A table showing 3 rows that summarize the findings of this paper, namely challenges in interdisciplinary research, usage and benefits of AI in interdisciplinary research, and challenges of using AI for interdisciplinary research.}
  \begin{tabular}[b]{l|ll}
    \textbf{Section} & \textbf{RQ} & \textbf{Findings} \\
    \midrule

    Section \ref{sec:rq1_results} & RQ1: Interdisciplinary challenges & \makecell[lc]{
        Searching for information in new disciplines. \\
        Aligning expectations when working across disciplines. \\
        Managing communication between communities.
    } \vspace{1mm}\\

    \arrayrulecolor{gray}\hline
    Section \ref{sec:rq2_results} & RQ2: Usage and benefits of AI & \makecell[lc]{
        Search engine with flexible queries. \\
        Translator between disciplines. \\
        In-discipline reviewer for preliminary feedback. \\
        Software development assistant.
    } \vspace{1mm}\\
    
    \arrayrulecolor{gray}\hline
    Section \ref{sec:rq3_results} & RQ3: Challenges of AI & \makecell[lc]{
        AI can cause an illusion of understanding when in unfamiliar disciplines. \\
        AI lacks the taste in research that comes with experience in a discipline. \\
        AI can reduce respect for disciplinary boundaries.
    } \vspace{1mm}\\
    
    \arrayrulecolor{black} \bottomrule
  \end{tabular}
\end{table*}

\subsection{RQ1: What challenges do researchers face when navigating a new discipline?}
\label{sec:rq1_results}

A few participants mentioned that the magnitudes of disciplinary differences can vary and consequently the challenges can vary (P15, P17, P19, P20); P19 noted, ``\textit{if something is close enough to what I do, you've got enough background knowledge that with some research, you can do the [literature] review yourself, and figure out where something fits in\ldots{}or whether the data seem like they were collected well\ldots{}Other things are much tougher where if I've got a question about genetics\ldots{}I just need to ask a geneticist.}''
Overall, participants reported three major challenges unique to navigating new disciplines: searching for information (Section \ref{sec:challenge_search}), aligning expectations (Section \ref{sec:challenge_exp}), and managing communication between communities (Section \ref{sec:challenge_com}).

\subsubsection{Searching for information in new disciplines.}
\label{sec:challenge_search}
Participants mentioned that navigating new disciplines presented unique challenges for information search, including unfamiliar and inconsistent terminology (P5, P6, P10-12, P16-19, P21, P23, S3, S11) that often represented assumed background knowledge (P1-5, P7-9, P11, P13, P14, P17, P18, P22, P23).
As P5 explained, ``\textit{\ldots{}that's [having sufficient background] the hardest, and usually people accumulate it over years\ldots{}When switching fields, you miss that}''.
For example, many participants noted that disciplines can have different terms for the same concept (P1-3, P17, P18, P20, P22) and likewise, the same term to signify different concepts (P6, P8, P14, P18, P21, P22).
In addition to searching for new work in external disciplines (P2, P6, P9, P16, P18, P21, P23), some participants mentioned needing to actively maintain their own expertise (P1, P2, P5, P16):
\begin{quote}
    ``\textit{As I'm focusing on this new stuff in this other area, I'm letting the other stuff slide in my area\ldots{}so it's just constant keeping up.}'' - P16
\end{quote}

\subsubsection{Aligning expectations when working across disciplines.}
\label{sec:challenge_exp}
Participants noted that expectations about methodology and analysis could differ between disciplines (P1, P3-5, P7-9, P11, P15, P18, P20-23, S4, S15) and influence results (e.g., experimental results varying with standard error expectations) (from P5, echoed by P17, P21, S15).
Additionally, at a higher level, disciplines can differ in how they frame problems and what they consider novel (P13, P15, P17, P18):
\begin{quote}
    ``\textit{I had to read up a lot more in terms of computation, because they [collaborators] wanted me to give them a problem, like compute X, Y, and Z\dots{}I needed to think like a computational biologist, because de novo, biologists are not trained to think what can be computed and what cannot be computed.}'' - P15
\end{quote}
Given these differences, calibrating expectations and priorities among collaborators can form roadblocks in how a project is framed and executed (P7, P8, P16, P17, P21-23), especially given that collaborators can have limited availability (P3, P4, P6, P8, P9, P14, P16-19, P23).
Giving an example of an interdisciplinary meeting, P22 said:
\begin{quote}
    ``\textit{I approach it in a very CS way, where we have this list of agenda items, and we're gonna get through them, and we're gonna make decisions about each thing, and then the humanities collaborators are like, actually, let's pause and discuss and critique together why we're even discussing inter-annotator agreement, or something like that\ldots{}people value different things, and it was good to hash it out.}''
\end{quote}

\subsubsection{Managing communication between communities.}
\label{sec:challenge_com}
When engaging with academic communities outside of their own, a few participants mentioned considering how in-discipline colleagues would perceive them and their work, particularly in how they speak and write (P14, P16, P18, P22, P23).
One example is citing references that the discipline considers conventional:
\begin{quote}
    ``\textit{I've seen this with machine learning people doing earth science stuff, where I'm like, why are they citing that for that?\ldots{}similarly, I want whoever's reading my paper to be confident that I did rigorous work, and citing the appropriately canonical literature on certain things is a way of communicating that you did.}'' - P23
\end{quote}
Hurdles in communication can subsequently exacerbate a lack of engagement between disciplines, despite potential similarities in their research:
\begin{quote}
    ``\textit{I think education is hard because if you do anything in, say, HCI, you tend to just build it and ignore all the other fields that exist and the things that they've done. And they're also ignoring you, because they're probably like `you didn't read our literature, you don't write the papers in the way that we write it'.}'' - P3
\end{quote}

\subsection{RQ2: How and why do researchers use AI when crossing disciplinary boundaries?}
\label{sec:rq2_results}
Participants described using AI systems as a search engine with flexible queries, a two-way translator between disciplines, an in-discipline reviewer to provide preliminary feedback, and a software development assistant.
For all uses except software development, participants queried chatbot interfaces (e.g., ChatGPT, Claude, Gemini); for software development, participants employed a mix of chatbots and coding agents (e.g., Claude Code, Copilot, Codex).
Through these uses, AI facilitated disciplinary boundary-crossing by making information easier to access, understand, and communicate.

\subsubsection{Search engine with flexible queries.}
\label{sec:search}
Many participants appreciated the ability to informally describe an idea that they have in natural language and ask an AI system to supply the keyword corresponding to that idea in a certain discipline, essentially translating their idea into the lingo of a target discipline (P1, P3, P4, P6-11, P13-18, P20-23, S3, S7, S11, S12, S17, S18):
\begin{quote}
    ``\textit{If there's a method that does a specific thing,\ldots{}and you just don't know the method that'll do that thing, you can ask [AI], `is there a method that does this thing'.}'' - P23
\end{quote}
As P20 put it, the keywords supplied by the system then help to ``\textit{unlock a whole literature}''.

Participants also used AI systems' flexible search for navigating large amounts of new information (P1, P4, P5, P8, P9, P14, P16, P18, S3, S7, S8, S18).
Sometimes, this involved organizing information; as P16 described ``\textit{it's like, wow, there's an overwhelming amount of stuff here. I need it organized in some sort of way where I can put my next step forward.}''
However, it also involved probing at sentiments in a discipline: ``\textit{I use AI for getting a kind of mean sense of what is going on in a discourse\ldots{}what are the kind of table stakes}'' (P18).


\subsubsection{Discipline translator.}
\label{sec:translator}
Building on the use of AI to translate concepts into keywords for search, participants also noted using AI as a two-way translation system: rephrasing information from other disciplines into words they understand and vice versa (P1, P3, P4, P6-11, P13-18, P20-23, S5, S11, S16, S17).
As P6 described for the former:
\begin{quote}
    ``\textit{[AI] does a pretty good job of breaking those down and explaining it, especially if you prompt it: `I'm in computer science, and I'm trying to understand this health paper, but I don't really understand a lot of the medical side of things'.}''
\end{quote}

Several participants also mention using AI to prepare for conversations with collaborators (P1, P3-5, P7, P8, P11, P13-15, P17) and to review notes after a meeting (P6):
\begin{quote}
    ``\textit{If I'm about to have a call with a collaborator and I realize that I don't know enough about XYZ, usually it's quick to just retrieve some examples, you know, use a chatbot to get at it quickly.}'' - P17
\end{quote}

\subsubsection{In-discipline reviewer for preliminary feedback.}
\label{sec:use_reviewer}
Relatedly, most participants reported using AI to check how other disciplines would perceive their work.
In particular, several mentioned using AI to provide a preliminary review of their work from an in-discipline perspective to ``\textit{spot any holes that have not been filled}'' (quoted from P5, echoed by P4, P8, P9, P11-14, P19, P20, P22, P23):
\begin{quote}
    ``\textit{For example, I finished a related work section, and I was like, here's the first draft that I have --do you think there are other theories that I need to draw on?\ldots{}And then it'll give me `here are the three theories you might want to consider', and then I'll think about whether or not that makes sense.}'' - P20
\end{quote}
This also extended to aspects like checking if they covered everything necessary for a presentation; as P14 describes: ``\textit{AI helps you to sort of say `this is what an average person that would read this abstract would expect from this presentation'\ldots{}it's like just going back through that tedium of `did I check this box, did I check that box', and it really helps with that.}''

\subsubsection{Software development assistant}


Across the previously-described uses, participants reported using a mix of chatbots (e.g., Claude, ChatGPT, Copilot, Gemini), with some orientation towards interfaces specifically for searching scholarly papers (e.g, Scholar Labs, Anara, Elicit).
The only exception to this tool use was coding agents for software development (e.g., Claude Code, Codex, Cursor) (P5, P6, P8, P12, P18, P20, P22, P23).
Even then, several participants preferred to copy and paste code snippets from a separate chat window rather than using agents with autonomous or semi-autonomous execution because of the monetary cost, a lack of coding expertise or control, and dissatisfaction with the output (P3, P7, P9, P13, P19).

Regardless of the type of AI system they used, participants who used AI for software development found it useful for quickly prototyping code (P1, P3, P6-8, P12, P13, P15, P19, P20, P22, P23), navigating large or unfamiliar codebases (P1, P5), and searching for libraries or packages to use (P17, P22), sometimes from unfamiliar disciplines:
\begin{quote}
    ``\textit{There are rare cases where I think [coding agents] are useful. If I don't know a particular package that is used by epidemiologists or something, that's a case where a coding agent can help me just quickly figure it out and make use of it. But most of the time, it's just chatbots effectively.}'' - P17
\end{quote}

\subsection{RQ3: What do researchers perceive to be the challenges of using AI in interdisciplinary research?}
\label{sec:rq3_results}
Informed by their experiences with using AI to navigate unfamiliar disciplines, participants raised three primary concerns; specifically, they noted that AI can provide an illusion of understanding (Section \ref{sec:illusion}), lacks the taste and judgment that comes with experience in a discipline (Section \ref{sec:taste}), and can reduce respect for the helpful differences that exist between disciplines (Section \ref{sec:heterogeneity}).

\subsubsection{AI can provide an illusion of understanding.}
\label{sec:illusion}
Participants were wary that the information they saw could shape their perception of a discipline, exacerbated by their lack of knowledge and intuition in the discipline (P1-7, P9-13, P15-18, P20, S2, S6, S15):
\begin{quote}
    ``\textit{You don't know what you don't know, and so you can't really assess the degree to which the summary is credible.}'' - P17
\end{quote}
Participants felt that AI sometimes provided a sense of ``\textit{false bravado}'' (P14) about their understanding, often contrasting with experiences when they do have background knowledge (e.g., ``\textit{I know how to program really well, and so I can audit what it's telling me}'' - P3). 
As a result, all participants expressed concerns about incorrect or incomplete information from AI.
The most common reason was hallucinations, such as fake papers (P5, P7, P11, P19) and incorrect interpretations of source text (P2, P5, P8, P10, P19, S11).
However, a few participants also noted concerns about bias in what AI systems used as sources (P3, P5, P12, P16, P20, S10, S17); ``\textit{I'm sure there's people citation hacking already with LLMs, and so there is a fear of is this really important information?...I don't know how the model was trained or what's weighted.}'' (P3).
A few participants mentioned that AI systems sometimes did not attribute to sources in the first place, reducing their ability to compare generated output against source information (P2, P10, P11, P19).

Some participants tried to mitigate this by ensuring that the system has explicit access to sources that they consider reputable, directly uploading PDFs (P6) or personal notes (P4, P14), or enabling web search (P2) and paper database access (P2, P13).
Additionally, some mentioned using AI for more surface-level information rather than details (P4-6, P14):
\begin{quote}
    ``\textit{I used AI at the very beginning, it's mainly just researching through the web interface\ldots{}If you add this too much and consider the details, then it may get wrong, and if you add this to summarize the paper without giving it the PDF, then this may just make up a few more things.}'' - P5
\end{quote}

All participants described triangulating information from AI output with traditional, non-AI resources to mitigate effects of misinformation and bias.
Examples given by participants included considering AI search results with those from databases or standard search engines (e.g., Google Scholar) (P2, P4, P7, P10, P15, P18, P20), bibliometric information (e.g., citation network) (P3, P4, P11, P10, P13, P17, P18, P21, P23), collaborator opinions (P7, P9, P10, P14, P16-18, P20, P22, S10), and experiences (e.g., museum field visits) (P7, P11):
\begin{quote}
    ``\textit{[AI] kind of gets diluted by the other things or sources that I'm relying on\ldots{}I do believe that it will probably focus on the most trendy things or whatnot\ldots{}but I think from my experience, because it's just one of the things that I'm using to collect information, it tends to get mitigated that way.}'' - P9
\end{quote}
A few participants also described comparing responses from different AI models to increase the likelihood of obtaining good information or ascertain whether or not they should trust the information (P3, P10, P12, P19):
\begin{quote}
    ``\textit{There was one case where I had this model [meaning a model the participant was constructing]\ldots{}and I gave it to Claude, Copilot, and I also have some local LLMs installed in LM Studio. I'm like, let's give it this code, and I'm gonna ask it `Are there any ways that I could consolidate or simplify or speed up this code?'\ldots{}A couple of them weren't very good, but more than one of the models gave suggestions on `here's this way that you could actually consolidate that, and here's some redundant steps in the flow of the code'.}'' - P19
\end{quote}

A complementary concern to incorrect information was incomplete information.
Some participants felt that their prompts or keywords would unduly influence responses (P2, P3, P7, P9-11, P15, P20); ``\textit{I think even saying the word `time gap' influences [the system] more towards `maybe this person is using more technical terminology, so they are probably interested in getting CS, or science'}'' (P3).
Participants also had concerns about what was lost when they relied on a system that had access to only partial information.
For example, P10 and P17, who work with metallurgy and medicine respectively, noted that practitioners' and stakeholders' opinions and experiences were not always represented in academic papers:
\begin{quote}
    ``\textit{\ldots{}there's stuff from the patient care perspective practitioners that is not encoded in papers\ldots{}So whenever anything interacts with other institutions that are non-academic, I find that there's a lot of implicit context that doesn't necessarily get cited or clearly stated in papers.}'' - P17
\end{quote}
Participants also gave examples of paywall restrictions (P6, P20) and as a more general limitation of digital tools, physical objects that cannot be fully digitized, such as ``\textit{glazed earthenware}'' (P11), ``\textit{microbial strains}'' (P15), and ``\textit{artifacts and soil samples}'' (P19):
\begin{quote}
    ``\textit{in art history, there are other contemporary art pieces that you're supposed to touch\ldots{}if it's a glazed earthenware, in picture, it might look identical to a porcelain. But, if you compare them in reality, you see that the latter one looks more translucent. I think there are a lot of subtle things that are missing during digitization.}'' - P11
\end{quote}
Lastly, a few participants noted that AI responses lose nuanced information that comes from reading between the lines, such as implied meanings (P1, P18), gaps (P5), and signals about credibility (P17, P18, P21):
\begin{quote}
    ``\textit{I'm often interested in how people think, not just what they think\ldots{}anything that summarizes or paraphrases the original text will lose some of that.}'' - P17
\end{quote}

\subsubsection{AI lacks the taste that comes with personal experience in a discipline.}
\label{sec:taste}

Almost all participants remarked that AI systems do not supply research taste or judgment (P1-4, P6-18, P20-23, S1).
Participants noted that AI may find information that is relevant but violates ``\textit{unspoken rules of the discipline}'' (P22-S).
For example, several participants noted that AI cannot provide information about the reputation or quality of a work or technique (P6, P7-9, P14, P18, P20, S1, S14, S15):
\begin{quote}
    ``\textit{It is that you just need to kind of know by osmosis what other people are thinking in the kind of nebulously-defined community.}'' - P18
\end{quote}
A few participants also described situations where AI made suggestions that deviated from discipline norms or sometimes feasibility (e.g., an infeasible research plan) (from P4, echoed by P8, S7).

Most participants preferred to retain oversight over deciding how to use information from AI systems or making downstream connections from one discipline to another, rather than letting the system decide (P1-3, P5, P7-14, P16-20, P22, P23, S18); as P1 remarked, ``\textit{I don't rely on the AI to make the decision about what's valuable. I rely on the AI to help me access the texts and work through them faster and more efficiently.}''
P18 noted that AI systems tended to simply accept work at face value:
\begin{quote}
    ``\textit{If you are reading a paper, you have certain expectations for what people do in a paper that make up a standard of evidence\ldots{}Whereas AI is very credulous about everything. If someone says `we present a novel, computationally efficient method', even if this is the least efficient thing you've ever seen, AI is not verifying that\ldots{}AI is just like, `you know what, you're right, that is novel'.}''
\end{quote}
For downstream connections in particular, several participants noted the importance of their personal experiences (P7, P8, P10, P11, P21-23), and relatedly, many mentioned that they value engaging with source information as a way of developing their own understanding of a discipline (P1, P2, P4, P8-10, P15, P17, P21):
\begin{quote}
    ``\textit{I'm not just developing my understanding of the text as an objective thing, I'm developing my personal reading and interpretation of the text as it fits into my trajectory as a researcher and my context and things.}'' - P1
\end{quote}

Many participants relied on collaborators to provide research judgment because of their membership in the discipline (P9, P12-14, P18, P20-23), consciousness of their positionality and limitations (P16, P18), and ability to ``\textit{push back}'' (P2) on ideas (P2, P3, P6-14, P17, P18, P20).
Examples included identifying promising directions (P1-23), supplying keywords to search for (P1, P2, P4, P6, P7, P9, P12, P15, P16, P20), providing feedback (P2, P3, P6-14, P17, P18, P20), connecting them with peers and opportunities in the discipline (P4, P12, P13, P16, P21-23), and sometimes directly handling project components (P13, P15, P22).
Participants contrast this with AI which they perceive as sycophantic (P2, P3, P5-7, P9, P11, P13, P15, P18, S1); as P9 noted, ``\textit{It's usually very affirmative, so I've never gotten anything that's super provoking that I'm like `yes, this is going to make the research so much better'. I think my advisor does a much better job at that.}''

\subsubsection{AI usage may reduce respect for disciplinary boundaries.}
\label{sec:heterogeneity}

Interestingly, some participants highlighted a possible risk in AI systems' ability to lower disciplinary boundaries: blurring the boundaries between disciplines too much could weaken the contributions of some.
P18 and P20 noted that this could reduce the ``\textit{helpful heterogeneity}'' (P20) that makes interdisciplinary research special and effective, particularly given the perception of AI as having a generic perspective (P10, P12, P14-16, P18, P20, P21, P23):
\begin{quote}
    ``\textit{If we already easily have access to everybody's disciplines, does that kind of interfere with how different disciplines end up being?\ldots{}It makes me wonder whether we're, in some ways, diminishing differences or helpful heterogeneity between fields by having access to it immediately.}'' - P20
\end{quote}

Relatedly, P10 and P18 mentioned the notion of respecting the time and experience it takes to earn membership in a discipline, something that all participants stated made their collaborators valuable to work with:
In relation to AI, P18 noted:
\begin{quote}
    ``\textit{It used to be that to understand those conventions meant that you had to spend a certain amount of time being in the field\ldots{}I feel like AI\ldots{}limits the degree to which you need to spend effort to get a sense of the norms of a field, but it also removes that sense of being changed by being in the field\ldots{}it also cheapens membership in the field, in that it now means less that you have committed to being a part of the field.}''
\end{quote}

For several participants, their position as an outsider of a discipline enabled them to provide a unique and novel perspective (e.g., ``\textit{working interdisciplinary sometimes makes you novel in nature}'') (quoted from P13, echoed by P10, P14, P15, P18, P20, P21, P23):
\begin{quote}
    ``\textit{Communicating in a way that this field is not used to was actually important\ldots{}We were aware that we were doing stuff where a lot of mining metallurgists are really going to care about the exact chemistry\ldots{}but that actually ended up being appreciated in peer review, where they were saying `oh, this is very novel and unique'.}'' - P10
\end{quote}
P18 described the outsider position as providing a useful ``\textit{partial perspective}'' because ``\textit{you can have different ideas if you are reading different things}'' from others in the discipline, suggesting that strictly conforming to a discipline is not always desired or beneficial.

\section{Discussion}

While AI applications in research are growing at a rapid rate, less attention has been given to how AI can support interdisciplinary research and the unique challenges that come with it.
Through our survey and interviews, we explored how interdisciplinary researchers used---and perceived the use of---AI when navigating unfamiliar disciplines.
Participants found AI helpful for searching with freeform queries, translating information to and from unfamiliar disciplines, and obtaining an idea of how someone from another discipline would perceive their work.
However, participants also described several challenges, specifically that AI can provide a false sense of understanding, lacks the taste and judgment that comes with discipline experience, and can reduce the heterogeneity that upholds interdisciplinary research.
In this section, we reflect on the role that AI can play in interdisciplinary research and design implications for such systems. We present three major takeaways, oriented to how AI systems should enable triangulation with alternative resources unique to a discipline (Section \ref{subsec:traingulate}), scaffold learning between disciplines (Section \ref{subsec:scaffold}), and support collaboration over individual execution (Section \ref{subsec:boundaries}).  

\subsection{AI systems should enable triangulation between resources.}
\label{subsec:traingulate}
Between the survey results and qualitative findings, participants found AI useful for navigating information in unfamiliar disciplines (Sections \ref{sec:survey_results}, \ref{sec:search}, and \ref{sec:translator}).
This usage resonates with \citet{liao2025llms}'s finding that the most common usage of AI among surveyed (in-discipline) researchers was information seeking.
However, compared to in-discipline researchers, interdisciplinary researchers often lack the intuition and taste that comes with years conducting research in a space. Participants in our studies found that this lack of intuition hampered their ability to adequately assess AI responses (e.g., determining the reputation of a paper or that a particular viewpoint represented a consensus). Generally, AI systems did not or could not provide this information (Section \ref{sec:taste}).
Participants addressed this challenge by cross-checking information from AI with traditional sources, such as source text and collaborators (Section \ref{sec:illusion}), though few affordances outside citations existed for them to do so easily.

AI tools for interdisciplinary researchers should incorporate support for triangulating generated output with other resources.
Mechanisms that display source text to a user can be especially helpful as a way to verify information and source quality (e.g., sharing cited papers with a collaborator).
Existing examples from AI-augmented reading support include tools that trace AI information to specific lines in source text \cite{Kambhamettu2025traceable,cohenwang2024contextcite} and forms of extractive summarization \cite{Gu2024airesilient}, such as highlighting important phrases from the source text directly. These designs could be adapted to specifically provide the intuition and taste-building necessary for interdisciplinary research, such as by retrieving context on a paper or examples from multiple papers discussing the same topic \citep{Dang2025CorpusStudioSE}.

Especially interesting was the fact that that many of our participants referenced resources beyond paper and text (e.g., podcasts, art). AI systems, enabled by more recent multi-modal models (i.e., models that can process vision, audio, and text \citep{caffagni-etal-2024-revolution}) could support searching and triangulating across these different media.

One consideration when researchers compare AI output to source information is mitigating verification fatigue (i.e., mental exhaustion from having to constantly validate information) \cite{Kim2025autoeth,hennigen2024symgen}.
In these situations, including scaffolding between the AI output and the source information (e.g., step-by-step explanations connecting the output to the source information) rather than leaving the user to figure out the connection could reduce the cognitive load of performing these checks.

\subsection{AI systems could scaffold learning between disciplines.}
\label{subsec:scaffold}
A recurrent finding through our studies was interdisciplinary researchers' desire to signal familiarity to other disciplines (e.g., finding canonical citations, Section \ref{sec:challenge_com}).
Participants often went beyond signaling by engaging with the discipline's source material to develop their background knowledge and intuition over time (Sections \ref{sec:taste}, \ref{sec:illusion}).
This aligns with prior work finding that interdisciplinary researchers often wanted to meaningfully learn from the disciplines with which they collaborated \cite{Lattuca_2002,Vanney_Mesurado_Sáenz_2024}.


Given participants' use of AI for translating between disciplines (Section \ref{sec:translator}), AI could supply learning opportunities rather than only translation support.
Building on our recommendations for triangulation, AI systems could provide scaffolding between generated output and the source information or between disciplines in a way that encourages and facilitates learning. For example, \citet{Chen2026scaffolding} explore the use of interactivity to improve scaffolding from reading static explanatory text to actively engaging with generated output. Systems could explain how a task is being executed rather than just executing it (e.g., mentioning conventional databases, flagship venues, and specific authors in the discipline while searching for literature). Although researchers may not always want or have the time to develop a detailed understanding of other disciplines, it could be available for researchers who envision engagement with a new discipline as a long-term goal.

\subsection{AI systems should support collaboration rather than automate boundary crossing.}
\label{subsec:boundaries}
AI can lower the barrier between disciplines by, for example, explaining concepts in new disciplines at various levels of detail, retrieving and explaining relevant papers, and acting as a sounding board from a particular discipline (Section \ref{sec:rq2_results}).
However, participants noted that this potential for AI could also reduce respect for the time and experience it takes to establish oneself in a new discipline (Section \ref{sec:heterogeneity}).
This echoes sentiments from prior work, particularly in what \citet{Ballantyne_2019} calls ``epistemic trespassing'' where experts from one discipline make claims for disciplines in which they have limited expertise. This exposes an interesting tension we uncovered in our study: while AI systems can enable interdisciplinary research by building bridges across disciplines, these same bridges might conversely lessen the perceived benefits of interdisciplinary research. 

This is where we see collaboration playing an important role.
Participants valued engaging with collaborators because collaborators' disciplinary membership meant that they had the training and exposure to provide background knowledge and intuition (Section \ref{sec:taste}).
In this way, collaboration enabled participants to reach across disciplinary boundaries without diminishing them.
Much of AI for science has focused on automating the research process or on accelerating individual scientists' workflows \cite{bran2024chemcrow,Savitsky_2025}.
However, rather than accelerating an individual researcher's ability to cross a disciplinary boundary, AI systems for interdisciplinary research should orient towards supporting collaboration between researchers.
One potential approach to this idea is the development of computational boundary objects.
Proposed by \citet{star1989boundary} in the context of scientific collaboration, boundary objects are representations that exist at the boundary of different disciplines and help mediate communication and understanding between those disciplines.
\citet{Wan2026knit} incorporated the idea of computational boundary objects with an AI-mediated framework (KNIT) that organizes inputs from different collaborators to surface misaligned expectations and understandings.
In terms of information exchange, some options for computational boundary objects could be AI-mediated collections of perspectives on individual objects (e.g., an art exhibit with commentaries from artists, historians, and museum curators where an AI model suggests points of convergence and divergence in their perspectives).
Outside of boundary objects, other approaches could leverage different forms of AI-mediated communication \cite{Hancock_Naaman_Levy_2020} (e.g., for collaborative ideation \cite{li2026GraftMind} and decision-making \cite{Lee2025mediate}).

\section{Limitations and Future Work}
\label{sec:limitations}

Our findings centered around information seeking and communication.
However, as several interview participants mentioned, interdisciplinary research can heavily depend on institutional structures and initiatives, such as tenure, time commitments, and infrastructure.
These aspects can significantly impact whether or not a researcher takes on interdisciplinary work in the first place \cite{Agapie2022hcihealth,Qin1998types}. We focused on information seeking and communication as this is where we saw the most potential for AI systems (as opposed to changing tenure review), though it limits the scope of our findings.


 Though recruitment messages were circulated across disciplines, respondents were primarily from STEM disciplines, with several from Computer Science, likely due to the authors' networks.
As a result of the latter, our respondent pool may have been biased towards being familiar with and inclined towards AI. Additionally, because all survey respondents had some existing research experience, most had established workflows prior to the increased availability of generative AI. This is in contrast to new researchers (e.g., undergraduates) who may have little experience with research tools before the generative AI era. Future work could explore how the resources they use and challenges they face differ, if at all.

Lastly, we explore this topic from individual respondents' perspectives rather than from team perspectives.
Although interview participants emphasized the importance of collaboration and provided insight into what collaboration is useful for, we did not directly observe team dynamics.
Team science is an interesting avenue for future work \cite{Anthony2023collab}, especially given recent work developing AI-mediated infrastructures for collaboration \cite{Wan2026knit,Cao2025medai}.







\section{Conclusion}

In this work, we present results from a survey of $62$ interdisciplinary researchers along with interviews with $23$ respondents.
Our findings describe the benefits and challenges that these researchers encounter when using AI to navigate new disciplines.
These findings motivate the development of AI systems that orient towards supporting collaboration rather than automating boundary crossing, scaffolding learning opportunities, and enabling users to triangulate information between different sources.
Future work can explore team-based perspectives and incoming researchers to expand our understanding of how to design AI systems that support interdisciplinary researchers across different stages and contexts.

\bibliographystyle{ACM-Reference-Format}
\bibliography{references,ARRreferences}

@String{Computing = "Computing" }

@String{Computer = "{IEEE} Computer" }

@String{Academic = "Academic Press" }

@String{Springer = "Springer-Verlag" }

@inproceedings{liao2025llms,
  author = {Liao, Zhehui and Antoniak, Maria and Cheong, Inyoung and Cheng, Evie Yu-Yen and Lee, Ai-Heng and Lo, Kyle and Chang, Joseph Chee and Zhang, Amy X},
  booktitle = {Proceedings of the 2nd Conference on Language Modeling (COLM)},
  month = oct,
  title = {{LLM}s as Research Tools: A Large Scale Survey of Researchers{'} Usage and Perceptions},
  url = {https://arxiv.org/abs/2411.05025},
  year = {2025}
}

@article{mudd2025screen,
author = {Mudd, Alexandra and Conroy, Tiffany and Voldbjerg, Siri Lygum and Goldschmied, Anita and Feo, Rebecca and Schuwirth, Lambert},
title = {Developing and Evaluating the Use of ChatGPT as a Screening Tool for Nurses Conducting Structured Literature Reviews: Proof of Concept Study Results},
journal = {Journal of Clinical Nursing},
year = {2025},
pages = {1-13},
doi = {https://doi.org/10.1111/jocn.17818},
url = {https://onlinelibrary.wiley.com/doi/abs/10.1111/jocn.17818},
eprint = {https://onlinelibrary.wiley.com/doi/pdf/10.1111/jocn.17818}
}

@misc{ChatGPTDeepResearch,
  author = {{OpenAI}},
  title = {{ChatGPT with Deep Research}},
  year = {2025},
  url = {https://chat.openai.com/}
}

@inproceedings{fok2023qlarify,
author = {Fok, Raymond and Chang, Joseph Chee and August, Tal and Zhang, Amy X. and Weld, Daniel S.},
title = {Qlarify: Recursively Expandable Abstracts for Dynamic Information Retrieval over Scientific Papers},
year = {2024},
isbn = {9798400706288},
publisher = {Association for Computing Machinery},
address = {New York, NY, USA},
url = {https://doi.org/10.1145/3654777.3676397},
doi = {10.1145/3654777.3676397},
booktitle = {Proceedings of the 37th Annual ACM Symposium on User Interface Software and Technology},
articleno = {145},
numpages = {21},
location = {Pittsburgh, PA, USA},
series = {UIST '24}
}

@article{rust2025lifestyle,
	title = {Evaluation of a large language model to simplify discharge summaries and provide cardiological lifestyle recommendations},
	volume = {5},
	issn = {2730-664X},
	url = {https://doi.org/10.1038/s43856-025-00927-2},
	doi = {10.1038/s43856-025-00927-2},
	number = {1},
	journal = {Communications Medicine},
	author = {Rust, Paul and Frings, Julian and Meister, Sven and Fehring, Leonard},
	month = may,
	year = {2025},
	pages = {208}
}

@inproceedings{Fok2023scim,
author = {Fok, Raymond and Kambhamettu, Hita and Soldaini, Luca and Bragg, Jonathan and Lo, Kyle and Hearst, Marti and Head, Andrew and Weld, Daniel S},
title = {Scim: Intelligent Skimming Support for Scientific Papers},
year = {2023},
isbn = {9798400701061},
publisher = {Association for Computing Machinery},
address = {New York, NY, USA},
url = {https://doi.org/10.1145/3581641.3584034},
doi = {10.1145/3581641.3584034},
booktitle = {Proceedings of the 28th International Conference on Intelligent User Interfaces},
pages = {476–490},
numpages = {15},
location = {Sydney, NSW, Australia},
series = {IUI '23}
}

@article{August2023paperplain,
author = {August, Tal and Wang, Lucy Lu and Bragg, Jonathan and Hearst, Marti A. and Head, Andrew and Lo, Kyle},
title = {
Paper Plain: Making Medical Research Papers Approachable to Healthcare Consumers with Natural Language Processing},
year = {2023},
issue_date = {October 2023},
publisher = {Association for Computing Machinery},
address = {New York, NY, USA},
volume = {30},
number = {5},
issn = {1073-0516},
url = {https://doi.org/10.1145/3589955},
doi = {10.1145/3589955},
journal = {ACM Trans. Comput.-Hum. Interact.},
month = sep,
articleno = {74},
numpages = {38}
}

@inproceedings{singh2025scholarqa,
    title = "Ai2 Scholar {QA}: Organized Literature Synthesis with Attribution",
    author = "Singh, Amanpreet  and
      Chang, Joseph Chee  and
      Haddad, Dany  and
      Naik, Aakanksha  and
      Hwang, Jena D.  and
      Kinney, Rodney  and
      Weld, Daniel S  and
      Downey, Doug  and
      Feldman, Sergey",
    editor = "Mishra, Pushkar  and
      Muresan, Smaranda  and
      Yu, Tao",
    booktitle = "Proceedings of the 63rd Annual Meeting of the Association for Computational Linguistics (Volume 3: System Demonstrations)",
    month = jul,
    year = "2025",
    address = "Vienna, Austria",
    publisher = "Association for Computational Linguistics",
    url = "https://aclanthology.org/2025.acl-demo.49/",
    doi = "10.18653/v1/2025.acl-demo.49",
    pages = "513--523",
    ISBN = "979-8-89176-253-4"
}

@Article{Rust2025llmsummary,
author={Rust, Paul
and Frings, Julian
and Meister, Sven
and Fehring, Leonard},
title={Evaluation of a large language model to simplify discharge summaries and provide cardiological lifestyle recommendations},
journal={Communications Medicine},
year={2025},
month=may,
day={29},
volume={5},
number={1},
pages={208},
issn={2730-664X},
doi={10.1038/s43856-025-00927-2},
url={https://doi.org/10.1038/s43856-025-00927-2}
}

@article{Whitfield2023elicit,
author = {Sharon Whitfield and Melissa A. Hofmann},
title = {Elicit: AI literature review research assistant},
journal = {Public Services Quarterly},
volume = {19},
number = {3},
pages = {201--207},
year = {2023},
publisher = {Routledge},
doi = {10.1080/15228959.2023.2224125},
URL = { 
        https://doi.org/10.1080/15228959.2023.2224125
},
eprint = { 
        https://doi.org/10.1080/15228959.2023.2224125
}
}

@inproceedings{Sharma2024Echo,
author = {Sharma, Nikhil and Liao, Q. Vera and Xiao, Ziang},
title = {Generative Echo Chamber? Effect of LLM-Powered Search Systems on Diverse Information Seeking},
year = {2024},
isbn = {9798400703300},
publisher = {Association for Computing Machinery},
address = {New York, NY, USA},
url = {https://doi.org/10.1145/3613904.3642459},
doi = {10.1145/3613904.3642459},
booktitle = {Proceedings of the 2024 CHI Conference on Human Factors in Computing Systems},
articleno = {1033},
numpages = {17},
location = {Honolulu, HI, USA},
series = {CHI '24}
}

@article{palmer1999interdisciplinary,
author = {Palmer, Carole L.},
title = {Structures and strategies of interdisciplinary science},
journal = {Journal of the American Society for Information Science},
volume = {50},
number = {3},
pages = {242-253},
doi = {https://doi.org/10.1002/(SICI)1097-4571(1999)50:3<242::AID-ASI7>3.0.CO;2-7},
year = {1999}
}

@article{Kapania2025hcillm,
author = {Kapania, Shivani and Wang, Ruiyi and Li, Toby Jia-Jun and Li, Tianshi and Shen, Hong},
title = { \'I\'m Categorizing LLM as a Productivity Tool\': Examining Ethics of LLM Use in HCI Research Practices},
year = {2025},
issue_date = {May 2025},
publisher = {Association for Computing Machinery},
address = {New York, NY, USA},
volume = {9},
number = {2},
url = {https://doi.org/10.1145/3711000},
doi = {10.1145/3711000},
journal = {Proc. ACM Hum.-Comput. Interact.},
month = may,
articleno = {CSCW102},
numpages = {26}
}

@inproceedings{guo2024personalized,
    title = "Personalized Jargon Identification for Enhanced Interdisciplinary Communication",
    author = "Guo, Yue  and
      Chang, Joseph Chee  and
      Antoniak, Maria  and
      Bransom, Erin  and
      Cohen, Trevor  and
      Wang, Lucy  and
      August, Tal",
    editor = "Duh, Kevin  and
      Gomez, Helena  and
      Bethard, Steven",
    booktitle = "Proceedings of the 2024 Conference of the North American Chapter of the Association for Computational Linguistics: Human Language Technologies (Volume 1: Long Papers)",
    month = jun,
    year = "2024",
    address = "Mexico City, Mexico",
    publisher = "Association for Computational Linguistics",
    url = "https://aclanthology.org/2024.naacl-long.255/",
    doi = "10.18653/v1/2024.naacl-long.255",
    pages = "4535--4550",
}

@inproceedings{Zheng2024disciplink,
author = {Zheng, Chengbo and Zhang, Yuanhao and Huang, Zeyu and Shi, Chuhan and Xu, Minrui and Ma, Xiaojuan},
title = {DiscipLink: Unfolding Interdisciplinary Information Seeking Process via Human-AI Co-Exploration},
year = {2024},
isbn = {9798400706288},
publisher = {Association for Computing Machinery},
address = {New York, NY, USA},
url = {https://doi.org/10.1145/3654777.3676366},
doi = {10.1145/3654777.3676366},
booktitle = {Proceedings of the 37th Annual ACM Symposium on User Interface Software and Technology},
articleno = {91},
numpages = {20},
location = {Pittsburgh, PA, USA},
series = {UIST '24}
}

@article{foster2004nonlinear,
author = {Foster, Allen},
title = {A nonlinear model of information-seeking behavior},
journal = {Journal of the American Society for Information Science and Technology},
volume = {55},
number = {3},
pages = {228-237},
doi = {https://doi.org/10.1002/asi.10359},
url = {https://asistdl.onlinelibrary.wiley.com/doi/abs/10.1002/asi.10359},
eprint = {https://asistdl.onlinelibrary.wiley.com/doi/pdf/10.1002/asi.10359},
year = {2004}
}

@inproceedings{Cao2025medai,
author = {Cao, Chen and Wu, Yu and Fang, Xiao Zoe and Liang, Zhenwen and Mamykina, Lena and Sbaffi, Laura and Xu, Xuhai},
title = {MedAI-SciTS: Enhancing Interdisciplinary Collaboration between AI Researchers and Medical Experts},
year = {2025},
isbn = {9798400713941},
publisher = {Association for Computing Machinery},
address = {New York, NY, USA},
url = {https://doi.org/10.1145/3706598.3713926},
doi = {10.1145/3706598.3713926},
booktitle = {Proceedings of the 2025 CHI Conference on Human Factors in Computing Systems},
articleno = {537},
numpages = {24},
location = {
},
series = {CHI '25}
}

@article{Mcmanus2006AFF,
  title={A framework for understanding uncertainty and its mitigation and exploitation in complex systems},
  author={Hugh L.N. Mcmanus and Daniel E. Hastings},
  journal={IEEE Engineering Management Review},
  year={2006},
  volume={34},
  pages={81-81},
  url={https://api.semanticscholar.org/CorpusID:29982179}
}

@inproceedings{Bao2025Words,
author = {Bao, Calvin and Shiue, Yow-Ting and Carpuat, Marine and Chan, Joel},
title = {Words as Bridges: Exploring Computational Support for Cross-Disciplinary Translation Work},
year = {2025},
isbn = {9798400713064},
publisher = {Association for Computing Machinery},
address = {New York, NY, USA},
url = {https://doi.org/10.1145/3708359.3712110},
doi = {10.1145/3708359.3712110},
booktitle = {Proceedings of the 30th International Conference on Intelligent User Interfaces},
pages = {1598–1623},
numpages = {26},
location = {
},
series = {IUI '25}
}

@article{leahey2017prominent,
  title={Prominent but less productive: The impact of interdisciplinarity on scientists’ research},
  author={Leahey, Erin and Beckman, Christine M and Stanko, Taryn L},
  journal={Administrative Science Quarterly},
  volume={62},
  number={1},
  pages={105--139},
  year={2017},
  publisher={Sage Publications Sage CA: Los Angeles, CA}
}

@article{bates1989design,
  title={The design of browsing and berrypicking techniques for the online search interface},
  author={Bates, Marcia J},
  journal={Online review},
  volume={13},
  number={5},
  pages={407--424},
  year={1989},
  publisher={MCB UP Ltd}
}

@article{Morris2023ScientistsPO,
  title={Scientists' Perspectives on the Potential for Generative AI in their Fields},
  author={Meredith Ringel Morris},
  journal={ArXiv},
  year={2023},
  volume={abs/2304.01420},
  url={https://api.semanticscholar.org/CorpusID:257921814}
}

@inproceedings{lucy-etal-2023-words,
    title = "Words as Gatekeepers: Measuring Discipline-specific Terms and Meanings in Scholarly Publications",
    author = "Lucy, Li  and
      Dodge, Jesse  and
      Bamman, David  and
      Keith, Katherine",
    booktitle = "Findings of the Association for Computational Linguistics: ACL 2023",
    month = jul,
    year = "2023",
    address = "Toronto, Canada",
    publisher = "Association for Computational Linguistics",
    url = "https://aclanthology.org/2023.findings-acl.433",
    doi = "10.18653/v1/2023.findings-acl.433",
    pages = "6929--6947",
}

@inproceedings{bhatt2025borderlands,
    title = "Research Borderlands: Analysing Writing Across Research Cultures",
    author = "Bhatt, Shaily  and
      August, Tal  and
      Antoniak, Maria",
    editor = "Che, Wanxiang  and
      Nabende, Joyce  and
      Shutova, Ekaterina  and
      Pilehvar, Mohammad Taher",
    booktitle = "Proceedings of the 63rd Annual Meeting of the Association for Computational Linguistics (Volume 1: Long Papers)",
    month = jul,
    year = "2025",
    address = "Vienna, Austria",
    publisher = "Association for Computational Linguistics",
    url = "https://aclanthology.org/2025.acl-long.1272/",
    doi = "10.18653/v1/2025.acl-long.1272",
    pages = "26238--26266",
    ISBN = "979-8-89176-251-0"
}

@inproceedings{Wan2026knit,
author = {Wan, Echo and Yin, Carrie and Ito, Akira and Gao, Ziwei and Jia, Jasper and Taoka, Yuki and Saito, Shigeki and Sadek, Malak and Mougenot, C\'{e}line},
title = {KNIT: Computational Boundary Objects for Real-Time Convergence in Interdisciplinary Teams},
year = {2026},
isbn = {9798400722783},
publisher = {Association for Computing Machinery},
address = {New York, NY, USA},
url = {https://doi.org/10.1145/3772318.3791921},
doi = {10.1145/3772318.3791921},
booktitle = {Proceedings of the 2026 CHI Conference on Human Factors in Computing Systems},
articleno = {723},
numpages = {22},
location = {
},
series = {CHI '26}
}

@article{Agapie2022hcihealth,
author = {Agapie, Elena and Haldar, Shefali and Poblete, Sharmaine G.},
title = {Using HCI in Cross-Disciplinary Teams: A Case Study of Academic Collaboration in HCI-Health Teams in the US Using a Team Science Perspective},
year = {2022},
issue_date = {November 2022},
publisher = {Association for Computing Machinery},
address = {New York, NY, USA},
volume = {6},
number = {CSCW2},
url = {https://doi.org/10.1145/3555610},
doi = {10.1145/3555610},
journal = {Proc. ACM Hum.-Comput. Interact.},
month = nov,
articleno = {552},
numpages = {35}
}

@misc{torres2026aimeetsscienceresearch,
      title={When AI Meets Science: Research Diversity, Interdisciplinarity, Visibility, and Retractions across Disciplines in a Global Surge}, 
      author={Andrés F. Castro Torres and Joan Giner-Miguelez and Mercè Crosas},
      year={2026},
      eprint={2605.06033},
      archivePrefix={arXiv},
      primaryClass={cs.DL},
      url={https://arxiv.org/abs/2605.06033}, 
}

@inproceedings{Lee2015moca,
author = {Lee, Charlotte P. and Paine, Drew},
title = {From The Matrix to a Model of Coordinated Action (MoCA): A Conceptual Framework of and for CSCW},
year = {2015},
isbn = {9781450329224},
publisher = {Association for Computing Machinery},
address = {New York, NY, USA},
url = {https://doi.org/10.1145/2675133.2675161},
doi = {10.1145/2675133.2675161},
booktitle = {Proceedings of the 18th ACM Conference on Computer Supported Cooperative Work \& Social Computing},
pages = {179–194},
numpages = {16},
location = {Vancouver, BC, Canada},
series = {CSCW '15}
}

@article{Spanner2001bordercrossings, title={Border Crossings: Understanding the Cultural and Informational Dilemmas of Interdisciplinary Scholars}, volume={27}, ISSN={0099-1333}, DOI={10.1016/S0099-1333(01)00220-8}, number={5}, journal={The Journal of Academic Librarianship}, author={Spanner, Don}, year={2001}, month={Sept}, pages={352–360} }

@article{Newby2011unfamiliar, title={Entering Unfamiliar Territory: Building an Information Literacy Course for Graduate Students in Interdisciplinary Areas}, volume={50}, ISSN={1094-9054}, number={3}, journal={Reference \& User Services Quarterly}, publisher={American Library Association}, author={Newby, Jill}, year={2011}, pages={224–229} }

@article{Daniel2022ChallengesFI,
  title={Challenges facing interdisciplinary researchers: Findings from a professional development workshop},
  author={Kristy L Daniel and Myra McConnell and Anita Schuchardt and Melanie E. Peffer},
  journal={PLoS ONE},
  year={2022},
  volume={17},
  url={https://api.semanticscholar.org/CorpusID:248262202}
}

@article {Vilhena_2014,
author = {Daril A. Vilhena and Jacob G. Foster and Martin Rosvall and Jevin D. West and James Evans and Carl T. Bergstrom },
title = {Finding Cultural Holes: How Structure and Culture Diverge in Networks of Scholarly Communication},
journal = {Sociological Science},
volume = {1},
number = {15},
issn = {2330-6696},
pages = {221--23},
year = {2014},
}

@misc{yao2025interdisciplinaryteamsfailsystematic,
      title={Why Interdisciplinary Teams Fail: A Systematic Analysis With Activity Theory in Clinical AI Collaboration}, 
      author={Bingsheng Yao and Yao Du and Yue Fu and Xuhai Xu and Yanjun Gao and Hong Yu and Dakuo Wang},
      year={2025},
      eprint={2410.00174},
      archivePrefix={arXiv},
      primaryClass={cs.HC},
      url={https://arxiv.org/abs/2410.00174}, 
}

@article{Dang2025CorpusStudioSE,
  title={CorpusStudio: Surfacing Emergent Patterns In A Corpus Of Prior Work While Writing},
  author={Hai Dang and Chelse Swoopes and Daniel Buschek and Elena L. Glassman},
  journal={Proceedings of the 2025 CHI Conference on Human Factors in Computing Systems},
  year={2025},
  url={https://api.semanticscholar.org/CorpusID:277066596}
}

@article{Leahey_2018, title={The Perks and Perils of Interdisciplinary Research}, volume={26}, DOI={10.1017/S1062798718000261}, number={S2}, journal={European Review}, author={Leahey, Erin}, year={2018}, pages={S55–S67}}

@article{Wudarczyk2021BringingTR,
  title={Bringing Together Robotics, Neuroscience, and Psychology: Lessons Learned From an Interdisciplinary Project},
  author={Olga A. Wudarczyk and Murat Kirtay and Anna K. Kuhlen and Rasha Abdel Rahman and John-Dylan Haynes and Verena V. Hafner and Doris Pischedda},
  journal={Frontiers in Human Neuroscience},
  year={2021},
  volume={15},
  url={https://api.semanticscholar.org/CorpusID:232387364}
}

@article{Strober2006HabitsOT,
  title={Habits of the Mind: Challenges for Multidisciplinary Engagement},
  author={Myra H. Strober},
  journal={Social Epistemology},
  year={2006},
  volume={20},
  pages={315 - 331},
  url={https://api.semanticscholar.org/CorpusID:15709099}
}

@inproceedings{caffagni-etal-2024-revolution,
    title = "The Revolution of Multimodal Large Language Models: A Survey",
    author = "Caffagni, Davide  and
      Cocchi, Federico  and
      Barsellotti, Luca  and
      Moratelli, Nicholas  and
      Sarto, Sara  and
      Baraldi, Lorenzo  and
      Baraldi, Lorenzo  and
      Cornia, Marcella  and
      Cucchiara, Rita",
    editor = "Ku, Lun-Wei  and
      Martins, Andre  and
      Srikumar, Vivek",
    booktitle = "Findings of the Association for Computational Linguistics: ACL 2024",
    month = aug,
    year = "2024",
    address = "Bangkok, Thailand",
    publisher = "Association for Computational Linguistics",
    url = "https://aclanthology.org/2024.findings-acl.807/",
    doi = "10.18653/v1/2024.findings-acl.807",
    pages = "13590--13618"
}

@article{Choi2007MultidisciplinarityIA,
  title={Multidisciplinarity, interdisciplinarity, and transdisciplinarity in health research, services, education and policy: 2. Promotors, barriers, and strategies of enhancement.},
  author={Bernard C. K. Choi and Anita W. P. Pak},
  journal={Clinical and investigative medicine. Medecine clinique et experimentale},
  year={2007},
  volume={30 6},
  pages={
          E224-32
        },
  url={https://api.semanticscholar.org/CorpusID:211765}
}

@article{huutoniemi2010analyzing,
  title={Analyzing interdisciplinarity: Typology and indicators},
  author={Huutoniemi, Katri and Klein, Julie Thompson and Bruun, Henrik and Hukkinen, Janne},
  journal={Research policy},
  volume={39},
  number={1},
  pages={79--88},
  year={2010},
  publisher={Elsevier}
}

@inbook{Palmer2017inforesearch, title={Information Research on Interdisciplinarity}, ISBN={978-0-19-873352-2}, url={https://doi.org/10.1093/oxfordhb/9780198733522.013.35}, DOI={10.1093/oxfordhb/9780198733522.013.35}, booktitle={The Oxford Handbook of Interdisciplinarity}, publisher={Oxford University Press}, author={Palmer, Carole L. and Fenlon, Katrina}, editor={Frodeman, Robert}, year={2017}, month=jan }

@book{Klein1991interdisc, address={Detroit}, edition={3. print}, title={Interdisciplinarity: history, theory, and practice}, ISBN={978-0-8143-2088-4}, publisher={Wayne State Univ. Press}, author={Klein, Julie Thompson}, year={1991} }

@book{Palmer2001workatbound, address={Dordrecht}, title={Work at the Boundaries of Science}, rights={http://www.springer.com/tdm}, ISBN={978-90-481-5884-3}, url={http://link.springer.com/10.1007/978-94-015-9843-9}, DOI={10.1007/978-94-015-9843-9}, publisher={Springer Netherlands}, author={Palmer, Carole L.}, year={2001}, language={en} }

@INPROCEEDINGS{Kim2025autoeth,
  author={Kim, Hannah and Pond, Sergei L. Kosakovsky and MacNeil, Stephen},
  booktitle={2025 IEEE Frontiers in Education Conference (FIE)}, 
  title={Conversations Over Clicks: Impact of Chatbots on Information Search in Interdisciplinary Learning}, 
  year={2025},
  volume={},
  number={},
  pages={1-9},
  doi={10.1109/FIE63693.2025.11328556}}

@article{Cibrian2022Interventions, title={Interdisciplinary Tensions When Developing Digital Interventions Supporting Individuals With ADHD}, volume={4}, ISSN={2673-253X}, DOI={10.3389/fdgth.2022.876039}, journal={Frontiers in Digital Health}, author={Cibrian, Franceli L. and Monteiro, Elissa and Schuck, Sabrina E. B. and Nelson, Michele and Hayes, Gillian R. and Lakes, Kimberley D.}, year={2022}, month=may, pages={876039} }

@inproceedings{Mitchell2021woz,
author = {Mitchell, Elliot and Mamykina, Lena},
title = {From the Curtain to Kansas: Conducting Wizard-of-Oz Studies in the Wild},
year = {2021},
isbn = {9781450380959},
publisher = {Association for Computing Machinery},
address = {New York, NY, USA},
url = {https://doi.org/10.1145/3411763.3443446},
doi = {10.1145/3411763.3443446},
booktitle = {Extended Abstracts of the 2021 CHI Conference on Human Factors in Computing Systems},
articleno = {46},
numpages = {6},
location = {Yokohama, Japan},
series = {CHI EA '21}
}

@inproceedings{Colusso2019sciencemodel,
author = {Colusso, Lucas and Jones, Ridley and Munson, Sean A. and Hsieh, Gary},
title = {A Translational Science Model for HCI},
year = {2019},
isbn = {9781450359702},
publisher = {Association for Computing Machinery},
address = {New York, NY, USA},
url = {https://doi.org/10.1145/3290605.3300231},
doi = {10.1145/3290605.3300231},
booktitle = {Proceedings of the 2019 CHI Conference on Human Factors in Computing Systems},
pages = {1–13},
numpages = {13},
location = {Glasgow, Scotland Uk},
series = {CHI '19}
}

@inproceedings{wadden2022multivers,
    title = "{M}ulti{V}er{S}: Improving scientific claim verification with weak supervision and full-document context",
    author = "Wadden, David  and
      Lo, Kyle  and
      Wang, Lucy Lu  and
      Cohan, Arman  and
      Beltagy, Iz  and
      Hajishirzi, Hannaneh",
    editor = "Carpuat, Marine  and
      de Marneffe, Marie-Catherine  and
      Meza Ruiz, Ivan Vladimir",
    booktitle = "Findings of the Association for Computational Linguistics: NAACL 2022",
    month = jul,
    year = "2022",
    address = "Seattle, United States",
    publisher = "Association for Computational Linguistics",
    url = "https://aclanthology.org/2022.findings-naacl.6/",
    doi = "10.18653/v1/2022.findings-naacl.6",
    pages = "61--76"
}

@inproceedings{Kambhamettu2025traceable,
author = {Kambhamettu, Hita and Flores, Jamie and Head, Andrew},
title = {Traceable Texts and Their Effects: A Study of Summary-Source Links in AI-Generated Summaries},
year = {2025},
isbn = {9798400713958},
publisher = {Association for Computing Machinery},
address = {New York, NY, USA},
url = {https://doi.org/10.1145/3706599.3719830},
doi = {10.1145/3706599.3719830},
booktitle = {Proceedings of the Extended Abstracts of the CHI Conference on Human Factors in Computing Systems},
articleno = {538},
numpages = {7},
location = {
},
series = {CHI EA '25}
}

@inproceedings{Jia2026llmcheck,
author = {Jia, Chenyan and Gondimalla, Apoorva and Zhang, Angie and Joseph Mullings, David and Boltz, Alexander and Lee, Min Kyung},
title = {Beyond Accuracy: Experts See AI Fact-Checks as Accurate but Less Useful},
year = {2026},
isbn = {9798400722783},
publisher = {Association for Computing Machinery},
address = {New York, NY, USA},
url = {https://doi.org/10.1145/3772318.3791625},
doi = {10.1145/3772318.3791625},
booktitle = {Proceedings of the 2026 CHI Conference on Human Factors in Computing Systems},
articleno = {929},
numpages = {22},
location = {
},
series = {CHI '26}
}

@inproceedings{Gu2024aidataanalysis,
author = {Gu, Ken and Shang, Ruoxi and Althoff, Tim and Wang, Chenglong and Drucker, Steven M.},
title = {How Do Analysts Understand and Verify AI-Assisted Data Analyses?},
year = {2024},
isbn = {9798400703300},
publisher = {Association for Computing Machinery},
address = {New York, NY, USA},
url = {https://doi.org/10.1145/3613904.3642497},
doi = {10.1145/3613904.3642497},
booktitle = {Proceedings of the 2024 CHI Conference on Human Factors in Computing Systems},
articleno = {748},
numpages = {22},
location = {Honolulu, HI, USA},
series = {CHI '24}
}

@article{Chen2025serenspark, title={Serendipitous sparks: AI information encounter, cognitive flexibility, AI literacy, and university student creativity}, volume={16}, ISSN={1664-1078}, DOI={10.3389/fpsyg.2025.1623730}, journal={Frontiers in Psychology}, author={Chen, Xiaoyan and Xiao, Limin}, pages={1623730}, year={2025} }

@article{Storey2025genaiinfo, title={Generative Artificial Intelligence: Evolving Technology, Growing Societal Impact, and Opportunities for Information Systems Research}, volume={27}, ISSN={1572-9419}, DOI={10.1007/s10796-025-10581-7}, author={Storey, Veda C. and Yue, Wei Thoo and Zhao, J. Leon and Lukyanenko, Roman}, year={2025}, month=oct, pages={2081–2102}, language={en} }

@inproceedings{duck2025seren,
author = {D\"{u}ck, Moritz and Holter, Steffen and Chan, Robin Shing Moon and Sevastjanova, Rita and El-Assady, Mennatallah},
title = {Finding Needles in Document Haystacks: Augmenting Serendipitous Claim Retrieval Workflows},
year = {2025},
isbn = {9798400713941},
publisher = {Association for Computing Machinery},
address = {New York, NY, USA},
url = {https://doi.org/10.1145/3706598.3713715},
doi = {10.1145/3706598.3713715},
booktitle = {Proceedings of the 2025 CHI Conference on Human Factors in Computing Systems},
articleno = {1003},
numpages = {17},
location = {
},
series = {CHI '25}
}

@article{Barley2022linguistic, title={Exploring the Relationship Between Interdisciplinary Ties and Linguistic Familiarity Using Multilevel Network Analysis}, volume={49}, ISSN={0093-6502}, DOI={10.1177/0093650220926001}, number={1}, journal={Communication Research}, publisher={SAGE Publications Inc}, author={Barley, William C. and Dinh, Ly and Workman, Hallie and Fang, Chengyu}, year={2022}, month=feb, pages={33–60}, language={EN} }

@article{Anthony2023collab, title={“Collaborating” with AI: Taking a System View to Explore the Future of Work}, volume={34}, ISSN={1047-7039}, DOI={10.1287/orsc.2022.1651}, number={5}, journal={Organization Science}, publisher={INFORMS}, author={Anthony, Callen and Bechky, Beth A. and Fayard, Anne-Laure}, year={2023}, month={Sept}, pages={1672–1694} }

@article{Qin1998types,
author = {Qin, Jian and Lancaster, F. W. and Allen, Bryce},
title = {Types and levels of collaboration in interdisciplinary research in the sciences},
journal = {Journal of the American Society for Information Science},
volume = {48},
number = {10},
pages = {893-916},
doi = {https://doi.org/10.1002/(SICI)1097-4571(199710)48:10<893::AID-ASI5>3.0.CO;2-X},
year = {1997}
}

@article{Klein2008evalid, series={The Science of Team Science}, title={Evaluation of Interdisciplinary and Transdisciplinary Research: A Literature Review}, volume={35}, ISSN={0749-3797}, DOI={10.1016/j.amepre.2008.05.010}, number={2, Supplement}, journal={American Journal of Preventive Medicine}, author={Klein, Julie T.}, year={2008}, month=aug, pages={S116–S123}, collection={The Science of Team Science} }

@article{Rangan2024cryodrgn, title={CryoDRGN-ET: deep reconstructing generative networks for visualizing dynamic biomolecules inside cells}, volume={21}, rights={2024 The Author(s), under exclusive licence to Springer Nature America, Inc.}, ISSN={1548-7105}, DOI={10.1038/s41592-024-02340-4}, number={8}, journal={Nature Methods}, publisher={Nature Publishing Group}, author={Rangan, Ramya and Feathers, Ryan and Khavnekar, Sagar and Lerer, Adam and Johnston, Jake D. and Kelley, Ron and Obr, Martin and Kotecha, Abhay and Zhong, Ellen D.}, year={2024}, month=aug, pages={1537–1545}, language={en} }

@book{Karns2026env, title={Using artificial intelligence to automate and expedite the harmonization of environmental data}, url={https://meetingorganizer.copernicus.org/EGU26/EGU26-12357.html}, DOI={10.5194/egusphere-egu26-12357}, note={DOI: 10.5194/egusphere-egu26-12357}, number={EGU26-12357}, institution={Copernicus Meetings}, author={Karns, Tyler and Hagen, Cedric and Deshpande, Krutika and SanClements, Michael and Laney, Christine and Ruddell, Benjamin and Loescher, Henry and Swetnam, Tyson}, year={2026}, month=mar, language={en} }

@article{Graziosi2023classics, title={Machine Learning and the Future of Philology: A Case Study}, volume={153}, ISSN={2575-7199}, number={1}, journal={TAPA}, publisher={Johns Hopkins University Press}, author={Graziosi, Barbara and Haubold, Johannes and Cowen-Breen, Charlie and Brooks, Creston}, year={2023}, pages={253–284} }

@inproceedings{Monge2025cowrite,
author = {Monge Roffarello, Alberto and Cal\`{o}, Tommaso and Scibetta, Luca and De Russis, Luigi},
title = {Investigating How Computer Science Researchers Design Their Co-Writing Experiences With AI},
year = {2025},
isbn = {9798400713941},
publisher = {Association for Computing Machinery},
address = {New York, NY, USA},
url = {https://doi.org/10.1145/3706598.3713205},
doi = {10.1145/3706598.3713205},
booktitle = {Proceedings of the 2025 CHI Conference on Human Factors in Computing Systems},
articleno = {1215},
numpages = {17},
location = {
},
series = {CHI '25}
}

@article{Fecher2025llmimplications, title={Friend or foe? Exploring the implications of large language models on the science system}, volume={40}, ISSN={1435-5655}, DOI={10.1007/s00146-023-01791-1}, number={2}, journal={AI \& SOCIETY}, author={Fecher, Benedikt and Hebing, Marcel and Laufer, Melissa and Pohle, Jörg and Sofsky, Fabian}, year={2025}, month=feb, pages={447–459}, language={en} }

@article{Gruda2024writing, title={Three ways ChatGPT helps me in my academic writing}, rights={2024 Springer Nature Limited}, url={https://www.nature.com/articles/d41586-024-01042-3}, DOI={10.1038/d41586-024-01042-3}, journal={Nature}, publisher={Nature Publishing Group}, author={Gruda, Dritjon}, year={2024}, month=apr, language={en} }

@inproceedings{Jakesch2023cowriting,
author = {Jakesch, Maurice and Bhat, Advait and Buschek, Daniel and Zalmanson, Lior and Naaman, Mor},
title = {Co-Writing with Opinionated Language Models Affects Users’ Views},
year = {2023},
isbn = {9781450394215},
publisher = {Association for Computing Machinery},
address = {New York, NY, USA},
url = {https://doi.org/10.1145/3544548.3581196},
doi = {10.1145/3544548.3581196},
booktitle = {Proceedings of the 2023 CHI Conference on Human Factors in Computing Systems},
articleno = {111},
numpages = {15},
location = {Hamburg, Germany},
series = {CHI '23}
}

@inproceedings{kim2024notsure,
author = {Kim, Sunnie S. Y. and Liao, Q. Vera and Vorvoreanu, Mihaela and Ballard, Stephanie and Vaughan, Jennifer Wortman},
title = {"I'm Not Sure, But...": Examining the Impact of Large Language Models' Uncertainty Expression on User Reliance and Trust},
year = {2024},
isbn = {9798400704505},
publisher = {Association for Computing Machinery},
address = {New York, NY, USA},
url = {https://doi.org/10.1145/3630106.3658941},
doi = {10.1145/3630106.3658941},
booktitle = {Proceedings of the 2024 ACM Conference on Fairness, Accountability, and Transparency},
pages = {822–835},
numpages = {14},
location = {Rio de Janeiro, Brazil},
series = {FAccT '24}
}

@inproceedings{liu2026writing,
author = {Liu, Yijun and Gallagher, John and Sterman, Sarah and August, Tal},
title = {From Crafting Text to Crafting Thought: Grounding AI Writing Support to Writing Center Pedagogy},
year = {2026},
isbn = {9798400722783},
publisher = {Association for Computing Machinery},
address = {New York, NY, USA},
url = {https://doi.org/10.1145/3772318.3790292},
doi = {10.1145/3772318.3790292},
booktitle = {Proceedings of the 2026 CHI Conference on Human Factors in Computing Systems},
articleno = {81},
numpages = {36},
location = {
},
series = {CHI '26}
}

@inproceedings{Gu2024airesilient,
author = {Gu, Ziwei and Arawjo, Ian and Li, Kenneth and Kummerfeld, Jonathan K. and Glassman, Elena L.},
title = {An AI-Resilient Text Rendering Technique for Reading and Skimming Documents},
year = {2024},
isbn = {9798400703300},
publisher = {Association for Computing Machinery},
address = {New York, NY, USA},
url = {https://doi-org.proxy2.library.illinois.edu/10.1145/3613904.3642699},
doi = {10.1145/3613904.3642699},
booktitle = {Proceedings of the 2024 CHI Conference on Human Factors in Computing Systems},
articleno = {898},
numpages = {22},
location = {Honolulu, HI, USA},
series = {CHI '24}
}

@misc{motivation,
  author       = {Volle, Adam},
  title        = {laptop computer},
  year         = {2026},
  url          = {https://www.britannica.com/technology/laptop-computer},
  note         = {Accessed: 2026-09-04},
  publisher    = {Encyclopedia Britannica, inc.}
}

@article{McHugh2012diffkappa, title={Interrater reliability: the kappa statistic}, volume={22}, ISSN={1330-0962}, abstractNote={The kappa statistic is frequently used to test interrater reliability. The importance of rater reliability lies in the fact that it represents the extent to which the data collected in the study are correct representations of the variables measured. Measurement of the extent to which data collectors (raters) assign the same score to the same variable is called interrater reliability. While there have been a variety of methods to measure interrater reliability, traditionally it was measured as percent agreement, calculated as the number of agreement scores divided by the total number of scores. In 1960, Jacob Cohen critiqued use of percent agreement due to its inability to account for chance agreement. He introduced the Cohen’s kappa, developed to account for the possibility that raters actually guess on at least some variables due to uncertainty. Like most correlation statistics, the kappa can range from −1 to +1. While the kappa is one of the most commonly used statistics to test interrater reliability, it has limitations. Judgments about what level of kappa should be acceptable for health research are questioned. Cohen’s suggested interpretation may be too lenient for health related studies because it implies that a score as low as 0.41 might be acceptable. Kappa and percent agreement are compared, and levels for both kappa and percent agreement that should be demanded in healthcare studies are suggested.}, number={3}, journal={Biochemia Medica}, author={McHugh, Mary L.}, year={2012}, month=oct, pages={276–282} }

@article{star1989boundary,
author = {Susan Leigh Star and James R. Griesemer},
title ={Institutional Ecology, `Translations' and Boundary Objects: Amateurs and Professionals in Berkeley's Museum of Vertebrate Zoology, 1907-39},
journal = {Social Studies of Science},
volume = {19},
number = {3},
pages = {387-420},
year = {1989},
doi = {10.1177/030631289019003001},
URL = { 
        https://doi.org/10.1177/030631289019003001
},
eprint = { 
        https://doi.org/10.1177/030631289019003001
}
}

@article{Li2023interdisc,
author = {Li, Bing and Chen, Shiji and Larivi{\`e}re, Vincent},
title = {Interdisciplinarity affects the technological impact of scientific research},
year = {2023},
issue_date = {Dec 2023},
publisher = {Springer-Verlag},
address = {Berlin, Heidelberg},
volume = {128},
number = {12},
issn = {0138-9130},
url = {https://doi.org/10.1007/s11192-023-04846-8},
doi = {10.1007/s11192-023-04846-8},
journal = {Scientometrics},
month = oct,
pages = {6527–6559},
numpages = {33}
}

@article{Smye_Frangi_2021, title={Interdisciplinary research: shaping the healthcare of the future}, volume={8}, ISSN={2514-6645}, DOI={10.7861/fhj.2021-0025}, number={2}, journal={Future Healthcare Journal}, author={Smye, Stephen W and Frangi, Alejandro F}, year={2021}, month={July}, pages={e218–e223} }

@article{Liu_Rong_2026, title={Beyond citation-based metrics: Measuring interdisciplinarity via SBERT semantic embeddings and its heterogeneous effects on citation impact}, volume={21}, ISSN={1932-6203}, DOI={10.1371/journal.pone.0354129}, number={7}, journal={PLOS ONE}, publisher={Public Library of Science}, author={Liu, Lu and Rong, Yu}, year={2026}, month={July}, pages={e0354129}, language={en} }

@article{Braun_Clarke_2019, title={Reflecting on reflexive thematic analysis}, volume={11}, ISSN={2159-676X}, DOI={10.1080/2159676X.2019.1628806}, number={4}, journal={Qualitative Research in Sport, Exercise and Health}, publisher={Routledge}, author={Braun, Virginia and Clarke, Victoria}, year={2019}, month=aug, pages={589–597} }

@article{Braun_Clarke_2006_ta, title={Using thematic analysis in psychology}, volume={3}, ISSN={1478-0887}, DOI={10.1191/1478088706qp063oa}, number={2}, journal={Qualitative Research in Psychology}, publisher={Routledge}, author={Braun, Virginia and Clarke, Victoria}, year={2006}, month=jan, pages={77–101} }

@inproceedings{cohenwang2024contextcite,
title={ContextCite: Attributing Model Generation to Context},
author={Benjamin Cohen-Wang and Harshay Shah and Kristian Georgiev and Aleksander Madry},
booktitle={ICML 2024 Next Generation of AI Safety Workshop},
year={2024},
url={https://openreview.net/forum?id=U4d4BIhwSt}
}

@inproceedings{
hennigen2024symgen,
title={Towards Verifiable Text Generation with Symbolic References},
author={Lucas Torroba Hennigen and Zejiang Shen and Aniruddha Nrusimha and Bernhard Gapp and David Sontag and Yoon Kim},
booktitle={First Conference on Language Modeling},
year={2024},
url={https://openreview.net/forum?id=fib9qidCpY}
}

@inproceedings{Grudin1990hcihistory,
author = {Grudin, Jonathan},
title = {The computer reaches out: the historical continuity of interface design},
year = {1990},
isbn = {0201509326},
publisher = {Association for Computing Machinery},
address = {New York, NY, USA},
url = {https://doi.org/10.1145/97243.97284},
doi = {10.1145/97243.97284},
booktitle = {Proceedings of the SIGCHI Conference on Human Factors in Computing Systems},
pages = {261–268},
numpages = {8},
location = {Seattle, Washington, USA},
series = {CHI '90}
}

@article{Olson_Olson_2003_hcihistory, title={Human-computer interaction: psychological aspects of the human use of computing}, volume={54}, ISSN={0066-4308}, DOI={10.1146/annurev.psych.54.101601.145044}, journal={Annual Review of Psychology}, author={Olson, Gary M. and Olson, Judith S.}, year={2003}, pages={491–516}, language={eng} }

@article{Ballantyne_2019, title={Epistemic Trespassing}, volume={128}, ISSN={0026-4423}, DOI={10.1093/mind/fzx042}, abstractNote={Epistemic trespassers judge matters outside their field of expertise. Trespassing is ubiquitous in this age of interdisciplinary research and recognizing this will require us to be more intellectually modest.}, number={510}, journal={Mind}, author={Ballantyne, Nathan}, year={2019}, month=apr, pages={367–395} }

@inproceedings{Mittal_Blodgett_Liao_2026, title={Learning by Chatting? Investigating the Impact of Generative AI on Information Seeking and Learning}, url={http://arxiv.org/abs/2606.11669}, DOI={10.48550/arXiv.2606.11669}, note={arXiv:2606.11669 [cs.HC]}, publisher={arXiv}, author={Mittal, Shravika and Blodgett, Su Lin and Liao, Q. Vera}, year={2026}, month=aug, booktitle={{9th AAAI Conference on AI, Ethics, and Society}} }

@article{llama3modelcard,
title={Llama 3 Model Card},
author={AI@Meta},
year={2024},
url = {https://huggingface.co/meta-llama/Meta-Llama-3-70B}
}

@article{bran2024chemcrow, title={Augmenting large language models with chemistry tools}, volume={6}, rights={2024 The Author(s)}, ISSN={2522-5839}, DOI={10.1038/s42256-024-00832-8}, number={5}, journal={Nature Machine Intelligence}, publisher={Nature Publishing Group}, author={M. Bran, Andres and Cox, Sam and Schilter, Oliver and Baldassari, Carlo and White, Andrew D. and Schwaller, Philippe}, year={2024}, month=may, pages={525–535}, language={en} }

@article{Savitsky_2025, title={Exclusive: Start-up FutureHouse debuts powerful AI ‘reasoning model’ for science}, volume={642}, rights={2025 Springer Nature Limited}, ISSN={1476-4687}, DOI={10.1038/d41586-025-01753-1}, abstractNote={The model, called ether0, outperforms other advanced AIs at chemistry tasks and is a stepping stone towards automating the entire research pipeline.}, number={8068}, journal={Nature}, publisher={Nature Publishing Group}, author={Savitsky, Zack}, year={2025}, month={June}, pages={552–553}, language={en} }

@inproceedings{kambhamettu2026litpivot,
  author = {Kambhamettu, Hita and Mishra, Bhavana Dalvi and Head, Andrew and Bragg, Jonathan and Naik, Aakanksha and Chang, Joseph Chee and Siangliulue, Pao},
  title = {LitPivot: Developing Well-Situated Research Ideas Through Dynamic Contextualization and Critique within the Literature Landscape},
  booktitle = {Proceedings of the 39th Annual {ACM} Symposium on User Interface Software and Technology},
  series = {{UIST}},
  year = {2026},
  month = nov,
  eprint = {2604.02600},
  archivePrefix = {arXiv}
}

@article{Lattuca_2002, title={Learning Interdisciplinarity: Sociocultural Perspectives on Academic Work}, volume={73}, ISSN={0022-1546}, number={6}, journal={The Journal of Higher Education}, publisher={Ohio State University Press}, author={Lattuca, Lisa R.}, year={2002}, pages={711–739} }

@inproceedings{Chen2026scaffolding,
author = {Chen, Zixin and Li, Haotian and Liu, Zhe and Qu, Huamin and Xie, Xing},
title = {From Passive Consumption to Active Interaction: Exploring Interactive LLM Scaffolding to Support Learning Engagement},
year = {2026},
isbn = {9798400722813},
publisher = {Association for Computing Machinery},
address = {New York, NY, USA},
url = {https://doi.org/10.1145/3772363.3798931},
doi = {10.1145/3772363.3798931},
booktitle = {Proceedings of the Extended Abstracts of the 2026 CHI Conference on Human Factors in Computing Systems},
articleno = {324},
numpages = {6},
location = {
},
series = {CHI EA '26}
}

@article{Vanney_Mesurado_Sáenz_2024, title={Measuring qualities needed for interdisciplinary work: The Intellectual Virtues for Interdisciplinary Research Scale (IVIRS)}, volume={19}, ISSN={1932-6203}, DOI={10.1371/journal.pone.0312938}, number={11}, journal={PLOS ONE}, publisher={Public Library of Science}, author={Vanney, Claudia E. and Mesurado, Belén and Sáenz, J. Ignacio Aguinalde}, year={2024}, month=nov, pages={e0312938}, language={en} }

@inproceedings{li2026GraftMind,
author = {li, Jiale and Fang, Yixuan and He, Xiangyang and Fan, Mingming},
title = {GraftMind: Facilitating Group Ideation with AI-Mediated Idea Sharing},
year = {2026},
isbn = {9798400722783},
publisher = {Association for Computing Machinery},
address = {New York, NY, USA},
url = {https://doi.org/10.1145/3772318.3791388},
doi = {10.1145/3772318.3791388},
booktitle = {Proceedings of the 2026 CHI Conference on Human Factors in Computing Systems},
articleno = {299},
numpages = {18},
location = {
},
series = {CHI '26}
}

@inproceedings{Lee2025mediate, address={New York, NY, USA}, series={IUI ’25 Companion}, title={Amplifying Minority Voices: AI-Mediated Devil’s Advocate System for Inclusive Group Decision-Making}, ISBN={979-8-4007-1409-2}, url={https://dl.acm.org/doi/10.1145/3708557.3716334}, DOI={10.1145/3708557.3716334}, booktitle={Companion Proceedings of the 30th International Conference on Intelligent User Interfaces}, publisher={Association for Computing Machinery}, author={Lee, SooHwan and Kim, Mingyu and Hwang, Seoyeong and Kim, Dajung and Lee, Kyungho}, year={2025}, month=mar, pages={17–21}, collection={IUI ’25 Companion} }

@article{Hancock_Naaman_Levy_2020, title={AI-Mediated Communication: Definition, Research Agenda, and Ethical Considerations}, volume={25}, ISSN={1083-6101}, DOI={10.1093/jcmc/zmz022}, number={1}, journal={Journal of Computer-Mediated Communication}, author={Hancock, Jeffrey T and Naaman, Mor and Levy, Karen}, year={2020}, month=jan, pages={89–100} }

\appendix
\section*{Appendix}

\section{Interview Participant and Survey Respondent Background}
\label{sec:app_participant_background}
Table \ref{tab:interview_participant_background} lists interview participants' ID, occupation, and disciplines along with years of experience in each discipline.
Tables \ref{tab:respondents_disciplines} and \ref{tab:respondents_disciplines_cont} list respondents' disciplines by mention along with the average years of experience across respondents within each discipline.

Survey respondents' professional background was as follows:
\begin{itemize}
    \item \textbf{Highest degree earned}: Master’s degree (33.9\%), Bachelor’s degree (25.8\%), Doctorate degree (33.9\%), Some college (1.6\%), Prefer not to answer (1.6\%), High school graduate (1.6\%), Professional degree beyond bachelor’s degree (1.6\%)
    \item \textbf{Institution} (could choose more than one): University (100.0\%), Non-profit (1.6\%)
    \item \textbf{Occupation} (could list more than one): Ph.D. student (53.2\%: 25.7\% 4th year, 22.9\% 5th year, 20.0\% 3rd year, 11.4\% 6th year, 11.4\% 1st year, 8.6\% 2nd year), Postdoctoral Associate/Fellow (9.6\%), Assistant Professor (6.5\%), Associate Professor (3.2\%), Research Assistant (3.2\%), Researcher (3.2\%), Undergraduate (3.2\%), Assistant Research Scientist (1.6\%), Associate Director and Instructor in a Graduate Program (1.6\%), Educational Research Associate (1.6\%), Graduate student (1.6\%), Lab Manager (1.6\%), Master's student (1.6\%), ML Researcher / Engineer at [company redacted] (1.6\%), Professor (1.6\%), Research Assistant Professor (1.6\%), Research Intern (1.6\%), Senior Research Specialist (1.6\%), Veterinary Researcher (1.6\%)
\end{itemize}
Survey respondents' demographic background was as follows; demographic questions were optional:
\begin{itemize}
    \item \textbf{Age}: 18-24 (9.7\%), 25-34 (53.2\%), 35-44 (8.1\%), 45-54 (4.8\%), 55-64 (3.2\%), Not reported (1.6\%)
    \item \textbf{Gender}: Man (44.4\%), Woman (38.1\%), Not reported (12.7\%), Non-binary (3.2\%), Prefer not to disclose (1.6\%) 
    \item \textbf{Race/Ethnicity}: White/Caucasian (40.3\%), Asian (35.5\%), Not reported (14.5\%), Hispanic and Latino (6.5\%), Middle Eastern (3.2\%), Black or African American (1.6\%), Prefer to self-describe: Jewish (1.6\%)
\end{itemize}

\begin{table*}[h]
\sffamily
\small
  \caption{\textbf{Interview participant background.}}
  \label{tab:interview_participant_background}
  \Description{A table showing a row for each interview participant that contains their participant ID, occupation, discipline(s), and years of experience in each discipline.}
  \begin{tabular}[b]{lll}
    \textbf{Participant} & \textbf{Occupation} & \textbf{Discipline (Years of Experience)} \\
    \midrule

    P1 & Research Assistant & \parbox[lc]{8cm}{
                            Computer Science (Programming) (12) $\bullet$ Literary Studies / Critical Theory (5) $\bullet$ Writing Studies (2) $\bullet$ Ethnomethodology / Conversation Analysis / Linguistic Anthropology (2) $\bullet$ Human-Computer Interaction (4)
    } \vspace{0.5mm}\\
    \midrule

    P2 & PhD student, 6th year & \parbox[lc]{8cm}{
                            Human-Computer Interaction (7) $\bullet$ Law (3)
    } \vspace{0.5mm}\\
    \midrule

    P3 & PhD student, 4th year & \parbox[l]{8cm}{
                            Computer Science (14) $\bullet$ Learning Sciences (3)
    } \vspace{0.5mm}\\
    \midrule

    P4 & PhD student, 3rd year & \parbox[l]{8cm}{
                            Human-Computer Interaction (4)
    } \vspace{0.5mm}\\
    \midrule

    P5 & Postdoc Researcher & \parbox[l]{8cm}{
                            Scientific Software (7) $\bullet$ Material Science (Polymer) (5) $\bullet$ Biophysics (2)
    } \vspace{0.5mm}\\
    \midrule

    P6 & PhD student, 2nd year & \parbox[l]{8cm}{
                            Computer Science (7) $\bullet$ Health (2)
    } \vspace{0.5mm}\\
    \midrule

    P7 & PhD student, 4th year & \parbox[l]{8cm}{
                            Information Science (6) $\bullet$ Computer Science (12) $\bullet$ Smell and Taste (5) $\bullet$ Neuroscience (1) $\bullet$ Chemistry (1)
    } \vspace{0.5mm}\\
    \midrule

    P8 & PhD student, 5th year & \parbox[l]{8cm}{
                            Computer Science (8)
    } \vspace{0.5mm}\\
    \midrule

    P9 & PhD student, 1st year & \parbox[l]{8cm}{
                            Human-Computer Interaction (3) $\bullet$ Computational Social Science (2) $\bullet$ Computer Science Education (4) $\bullet$ Systems (2)
    } \vspace{0.5mm}\\
    \midrule

    P10 & Postdoc & \parbox[l]{8cm}{
                            Geology (9) $\bullet$ Policy (2) $\bullet$ Nuclear Weapons Non-proliferation (2)
    } \vspace{0.5mm}\\
    \midrule

    P11 & PhD student, 1st year & \parbox[l]{8cm}{
                            Museum Studies (4) $\bullet$ Art History (3) $\bullet$ Computer Science (2) $\bullet$ Information Science (1)
    } \vspace{0.5mm}\\
    \midrule

    P12 & PhD student, 3rd year & \parbox[l]{8cm}{
                            Informatics (3) $\bullet$ Architecture / Historic Preservation (9) $\bullet$ Landscape Architecture (4) $\bullet$ Architectural Engineering (8) $\bullet$ Extended Reality (XR) / Game Technologies (5)
    } \vspace{0.5mm}\\
    \midrule

    P13 & PhD student, 3rd year & \parbox[l]{8cm}{
                            Architecture (14) $\bullet$ Landscape Architecture (10) $\bullet$ Historic Preservation (8) $\bullet$ Game Studies (8) $\bullet$ Informatics (2)
    } \vspace{0.5mm}\\
    \midrule

    P14 & Educational Research Associate & \parbox[l]{8cm}{
                            Chemistry (20) $\bullet$ Education (10) $\bullet$ Education Research (6)
    } \vspace{0.5mm}\\
    \midrule

    P15 & Researcher & \parbox[l]{8cm}{
                            Chemistry (15) $\bullet$ Biochemistry (5) $\bullet$ Microbiology (2)
    } \vspace{0.5mm}\\
    \midrule

    P16 & Associate Professor & \parbox[l]{8cm}{
                            Graphic Design/Branding (28) $\bullet$ Education (Higher Ed.) (20) $\bullet$ UIUX Design (26)
    } \vspace{0.5mm}\\
    \midrule

    P17 & Assistant Professor & \parbox[l]{8cm}{
                            Artificial Intelligence (6) $\bullet$ Human-Computer Interaction (7) $\bullet$ Social/Behavioral Science (4) $\bullet$ Natural Language Processing (2) $\bullet$ Signal Processing (8)
    } \vspace{0.5mm}\\
    \midrule

    P18 & PhD Student, 6th year & \parbox[l]{8cm}{
                            Computer Science (15) $\bullet$ Natural Language Processing (8) $\bullet$ Science and Technology Studies (3) $\bullet$ Human Computer Interaction (1)
    } \vspace{0.5mm}\\
    \midrule

    P19 & Assistant Research Scientist & \parbox[l]{8cm}{
                            Archaeology (15)
    } \vspace{0.5mm}\\
    \midrule

    P20 & Assistant Professor & \parbox[l]{8cm}{
                            Human-Computer Interaction (8) $\bullet$ Communication (4) $\bullet$ Journalism (6) $\bullet$ Computational Social Science (6)
    } \vspace{0.5mm}\\
    \midrule
    
    P21 & Associate Professor & \parbox[l]{8cm}{
                            History (19) $\bullet$ Philosophy (10)
    } \vspace{0.5mm}\\
    \midrule
    
    P22 & Assistant Professor & \parbox[l]{8cm}{
                            Natural Language Processing (13) $\bullet$ Cultural Analytics (10) $\bullet$ Healthcare (13)
    } \vspace{0.5mm}\\
    \midrule

    P23 & Postdoctoral Associate & \parbox[l]{8cm}{
                            Atmospheric/Oceanic Science (5) $\bullet$ Machine Learning (5) $\bullet$ Environmental Engineering/Hydrology (7)
    } \vspace{0.5mm}\\
    
    \arrayrulecolor{black} \bottomrule
  \end{tabular}
\end{table*}

\begin{table*}[h]
\sffamily
\small
  \caption{\textbf{Respondent disciplines by mention.} ``Mentions'' represents how many respondents mentioned the discipline. ``Avg. Years'' shows the mean and standard deviation of years of experience for each discipline mentioned.}
  \label{tab:respondents_disciplines}
  \Description{A table listing each discipline and the number of respondents who indicated it along with the average years of experience with standard deviation.}
  \begin{tabular}[b]{lll}
    \textbf{Mentions} & \textbf{Discipline} & \textbf{Avg. Years} \\
    \midrule
     $18$ & Computer Science & $7.72 \pm 4.04$ \\
     $15$ & Human-Computer Interaction & $4.47 \pm 2.19$ \\
     $8$ & Artificial Intelligence / Machine Learning & $5.0 \pm 1.73$ \\
     $6$ & Education & $6.67 \pm 2.21$ \\
     $5$ & Chemistry & $10.4 \pm 7.89$ \\
     $4$ & Materials Science & $5.75 \pm 1.92$ \\
     $4$ & Natural Language Processing & $7.5 \pm 3.91$ \\
     $2$ & Finance & $1.5 \pm 0.5$ \\
     $2$ & English & $4.5 \pm 3.5$ \\
     $2$ & Health & $1.5 \pm 0.5$ \\
     $2$ & Learning Sciences & $3.0 \pm 0.0$ \\
     $2$ & Computational Social Science & $4.0 \pm 2.0$ \\
     $2$ & Computer Science Education & $4.5 \pm 0.5$ \\
     $2$ & Information Science & $3.5 \pm 2.5$ \\
     $2$ & Neuroscience & $3.0 \pm 2.0$ \\
     $2$ & Biophysics & $4.5 \pm 2.5$ \\
     $2$ & Mechanical Engineering & $6.5 \pm 2.5$ \\
     $2$ & Architecture & $11.5 \pm 2.5$ \\
     $2$ & Landscape Architecture & $7.0 \pm 3.0$ \\
     $2$ & Historic Preservation & $8.5 \pm 0.5$ \\
     $2$ & Extended Reality (XR) & $4.0 \pm 1.0$ \\
     $2$ & Policy & $3.0 \pm 1.0$ \\
     $2$ & Political Science & $2.5 \pm 0.5$ \\
     $2$ & Social Science & $2.5 \pm 0.5$ \\
     $2$ & Cultural Analytics & $6.0 \pm 4.0$ \\
     $2$ & Communication & $6.0 \pm 2.0$ \\
     $2$ & Mathematics & $7.0 \pm 3.0$ \\
     $1$ & Social/Behavioral Science & $3.0 \pm 1.0$ \\
     $1$ & Law & $3.0 \pm 0.0$ \\
     $1$ & Literary Studies / Critical Theory & $5.0 \pm 0.0$ \\
     $1$ & Writing Studies & $2.0 \pm 0.0$ \\
     $1$ & Ethnomethodology / Conversation Analysis / Linguistic Anthropology & $2.0 \pm 0.0$ \\
     $1$ & Software Engineering & $5.0 \pm 0.0$ \\
     $1$ & Biomedical Imaging & $2.0 \pm 0.0$ \\
     $1$ & Bioinformatics & $3.0 \pm 0.0$ \\
     $1$ & Systems (Computer Science) & $2.0 \pm 0.0$ \\
     $1$ & Psychology & $4.0 \pm 0.0$ \\
     $1$ & Biology & $4.0 \pm 0.0$ \\
     $1$ & Smell and Taste & $5.0 \pm 0.0$ \\
     $1$ & Scientific Software & $7.0 \pm 0.0$ \\
     $1$ & Cognitive Science & $30.0 \pm 0.0$ \\
     $1$ & Statistics & $15.0 \pm 0.0$ \\
     $1$ & Decision Sciences & $5.0 \pm 0.0$ \\
     $1$ & Materials Engineering & $2.0 \pm 0.0$ \\
     $1$ & Tissue Engineering & $2.0 \pm 0.0$ \\
     $1$ & Business & $2.0 \pm 0.0$ \\
     $1$ & Anthropology & $8.0 \pm 0.0$ \\
     $1$ & Design & $3.0 \pm 0.0$ \\
     $1$ & Behavioral Health & $3.0 \pm 0.0$ \\
     $1$ & Museum Studies & $4.0 \pm 0.0$ \\
    \bottomrule
  \end{tabular}
\end{table*}

\begin{table*}[h]
\sffamily
\small
  \caption{\textbf{Respondent disciplines by mention (cont.).} ``Mentions'' represents how many respondents mentioned the discipline. ``Avg. Years'' shows the mean and standard deviation of years of experience for each discipline mentioned.}
  \label{tab:respondents_disciplines_cont}
  \Description{A table listing each discipline and the number of respondents who indicated it along with the average years of experience with standard deviation, continued from the previous table.}
  \begin{tabular}[b]{lll}
    \textbf{Mentions} & \textbf{Discipline} & \textbf{Avg. Years} \\
    \midrule
     $1$ & Art History & $3.0 \pm 0.0$ \\
     $1$ & Game Studies & $8.0 \pm 0.0$ \\
     $1$ & Architectural Engineering & $8.0 \pm 0.0$ \\
     $1$ & Game Technologies & $5.0 \pm 0.0$ \\
     $1$ & Quantitative Biology & $3.0 \pm 0.0$ \\
     $1$ & Geology & $9.0 \pm 0.0$ \\
     $1$ & Nuclear Weapons Non-proliferation & $2.0 \pm 0.0$ \\
     $1$ & Atmospheric/Oceanic Science & $5.0 \pm 0.0$ \\
     $1$ & Environmental Engineering/Hydrology & $7.0 \pm 0.0$ \\
     $1$ & Haptics & $1.0 \pm 0.0$ \\
     $1$ & Plant Biology & $25.0 \pm 0.0$ \\
     $1$ & Security and Privacy & $10.0 \pm 0.0$ \\
     $1$ & Signal Processing & $8.0 \pm 0.0$ \\
     $1$ & Digital Humanities & $2.0 \pm 0.0$ \\
     $1$ & Science and Technology Studies & $3.0 \pm 0.0$ \\
     $1$ & Computational Neuroscience & $2.0 \pm 0.0$ \\
     $1$ & Digital Signal Processing & $1.0 \pm 0.0$ \\
     $1$ & Atmospheric Sciences & $11.0 \pm 0.0$ \\
     $1$ & Education Research & $6.0 \pm 0.0$ \\
     $1$ & Graphic Design/Branding & $28.0 \pm 0.0$ \\
     $1$ & Education (Higher Ed.) & $20.0 \pm 0.0$ \\
     $1$ & UIUX Design & $26.0 \pm 0.0$ \\
     $1$ & Global Studies & $10.0 \pm 0.0$ \\
     $1$ & International Educational Development & $22.0 \pm 0.0$ \\
     $1$ & Biochemistry & $5.0 \pm 0.0$ \\
     $1$ & Microbiology & $2.0 \pm 0.0$ \\
     $1$ & Computational Biology & $5.0 \pm 0.0$ \\
     $1$ & Social Media/Social Computing & $3.0 \pm 0.0$ \\
     $1$ & Archaeology & $15.0 \pm 0.0$ \\
     $1$ & History & $19.0 \pm 0.0$ \\
     $1$ & Philosophy & $10.0 \pm 0.0$ \\
     $1$ & Healthcare & $13.0 \pm 0.0$ \\
     $1$ & Economics & $6.0 \pm 0.0$ \\
     $1$ & Data Science & $6.0 \pm 0.0$ \\
     $1$ & Rehabilitation Counseling & $13.0 \pm 0.0$ \\
     $1$ & Special Education & $4.0 \pm 0.0$ \\
     $1$ & Educational Psychology & $7.0 \pm 0.0$ \\
     $1$ & Vocational Rehabilitation & $13.0 \pm 0.0$ \\
     $1$ & Journalism & $6.0 \pm 0.0$ \\
     $1$ & Information Systems Management & $5.0 \pm 0.0$ \\
     $1$ & Computational Economics/Finance & $4.0 \pm 0.0$ \\
     $1$ & Accessibility & $2.0 \pm 0.0$ \\
     $1$ & Creativity Support & $1.0 \pm 0.0$ \\
     $1$ & Veterinary Medicine & $34.0 \pm 0.0$ \\
     $1$ & Electrical Engineering & $28.0 \pm 0.0$ \\
     $1$ & Ultrasound Research & $28.0 \pm 0.0$ \\
     $1$ & Programming Languages & $3.0 \pm 0.0$ \\
     $1$ & Operating Systems & $4.0 \pm 0.0$ \\
     $1$ & AI Infrastructure & $2.0 \pm 0.0$ \\
     $1$ & Distributed Storage Systems & $2.0 \pm 0.0$ \\
    \bottomrule
  \end{tabular}
\end{table*}

\section{Survey Questions}
\label{sec:survey_questions}
The majority of questions were multiple-choice, using 5-point scales for rating questions.

\subsection{Background}
\noindent Q. List the discipline(s) that you work in and the number of years of experience in each (rounding up). You can list at most 5 disciplines.

\noindent Q. Please link at least 1 of your scholar profiles.
\begin{itemize}
    \item Google Scholar profile link:
    \item Semantic Scholar profile link:
    \item ORCID ID profile link:
    \item dblp profile link:
\end{itemize}

\noindent Q. (Optional) If it would be helpful for showing your research or projects, please provide a link to your personal website or department/company profile. [Free text entry]

\noindent Q. (Optional) Briefly describe any interdisciplinary research problem(s) you are working on. Bullet points are fine as long as there is sufficient description.

\noindent Q. What type of institution do you primarily conduct research in? (Select all that apply)
\begin{itemize}
    \item University
    \item Commercial Company
    \item Non-profit
    \item Governmental
    \item I do not conduct research in a professional capacity
    \item Other:
\end{itemize}

\noindent Q. What is the highest degree that you have received? (If you are currently enrolled in school, please indicate the highest degree you have received.)
\begin{itemize}
    \item Less than high school
    \item High school graduate
    \item Some college
    \item Associates degree (for example: AA, AS)
    \item Bachelor’s degree (for example: BA. BS)
    \item Master’s degree (for example: MA, MS, MEng, MEd, MSW, MBA)
    \item Professional degree beyond bachelor’s degree (for example: MD, DDS, DVM, LLB, JD)
    \item Doctorate degree (for example, PhD, EdD)
    \item Prefer not to answer
\end{itemize}

\noindent Q. What is your current position or occupation (e.g., Undergraduate, Engineer, Postdoc, Prefer not to answer)? [Free text entry]

\noindent Q. (Optional) If you are a PhD student, how many years have you completed?
\begin{itemize}
    \item I am not a PhD student
    \item 0
    \item 1
    \item 2
    \item 3
    \item 4
    \item 5
    \item 6
    \item 7+
\end{itemize}

\noindent Q. How familiar are you with Artificial Intelligence (AI) or AI-based tools such as ChatGPT, Gemini, and Claude?
\begin{itemize}
    \item Not familiar at all, I have never heard of them
    \item Slightly familiar, I have heard of them or have some idea of
    \item Moderately familiar, I know what they are and can explain
    \item Very familiar, I have technical knowledge of what they are and how they work
    \item Extremely familiar, I consider myself an expert on them
\end{itemize}

\noindent Q. How often do you use AI or AI-based tools?
\begin{itemize}
    \item Never
    \item Rarely
    \item Occasionally
    \item Regularly
    \item Frequently
\end{itemize}

\noindent Q. Overall, how do you feel about AI and AI-based tools?
\begin{itemize}
    \item Negative
    \item Somewhat negative
    \item Neutral
    \item Somewhat positive
    \item Positive
\end{itemize}

\noindent Q. What non-digital resources do you consult when exploring disciplines beyond your primary discipline and how often? [Respondent selects from 5-point Never to Frequently scale for each item]
\begin{itemize}
    \item Speaking to collaborators
    \item Attending seminars, conferences, or workshops
    \item Reading archives or physical books
\end{itemize}

\noindent Q. (Optional) If you use any non-digital resources that are not listed above, please list them here along with how frequently you use them. [Free text entry]

\subsection{Interdisciplinary Experiences}
\label{sec:interdisc_exp_questions}

\noindent Q. How often do you experience the following when exploring disciplines beyond your primary discipline? [Respondent selects from 5-point Never to Frequently scale for each item]

\noindent\underline{Navigating Information}
\begin{itemize}
    \item Unfamiliar terminology or jargon
    \item Deciding what to ask or search for
\end{itemize}
\noindent\underline{Assessing Information}
\begin{itemize}
    \item Considering a work’s context in its discipline
    \item Deciding when information is valuable
    \item Deciding when you have sufficient information
\end{itemize}
\noindent\underline{Connecting Across Disciplines}
\begin{itemize}
    \item Communication with people in other disciplines
    \item Different methodological expectations
    \item Different expectations surrounding analysis and evidence
    \item Different writing styles
\end{itemize}

\noindent Q. How challenging do you find each experience (when not using Artificial Intelligence (AI))? If you do not experience an item, please select "I do not experience this" in the corresponding row. [For the same items as in the previous question, respondent selects from ``I do not experience this'', ``Not challenging'', ``Slightly challenging'', ``Moderately challenging'', ``Very challenging'', and ``Extremely challenging'']

\noindent Q. Do you use AI for handling any of these experiences, and if so, how useful do you find it? If you do not use AI for an item, please select "I do not use AI for this" in the corresponding row. [For the same items as in the previous question, respondent selects from ``I do not use AI for this'', ``Not useful'', ``Slightly useful'', ``Moderately useful'', ``Very useful'', and ``Extremely useful'']

\noindent Q. (Optional) What benefits and/or challenges do you experience with AI when exploring disciplines beyond your primary discipline? [Free text entry]

\noindent Q. (Optional) If you experience anything when exploring disciplines beyond your primary discipline that was not included in the previous questions, please list them here along with how frequently you encounter them, how challenging you find them, and how useful you find AI for handling them. [Free text entry]

\subsection{Follow-up Questions}
\noindent Q. Would you be interested in participating in a follow-up interview? We are conducting the interviews to help contextualize and understand survey responses more deeply.

Details: We will select a few survey respondents to participate in a 1-hour interview over Zoom in which we will ask about your survey responses and, with your permission, may ask you to demonstrate some of your workflow (e.g., how you search for concepts, how you read a paper). Interview participants will be compensated with a \$15 Amazon gift card.

You may view the consent form for the interview here: [PDF link]. Note that you are not providing your consent here; we would contact you using the email you provided earlier to provide the consent form and schedule the interview.

\begin{itemize}
    \item Yes, I'm interested in participating
    \item No, I'm not interested in participating
\end{itemize}

\noindent Q. Would you like to be entered in the raffle? We will randomly select 5 respondents to each receive a \$20 Amazon gift card. If you are selected, we will contact you using the email you provided earlier to first verify it and then send the gift card.
\begin{itemize}
    \item Yes
    \item No
\end{itemize}

\noindent Q. (Optional) Is there anything else that you would like the research team to know? You can also email [research team email] if you have any questions or comments after submitting. [Free text entry]

\subsection{Optional Demographics}
\noindent Q. (Optional) What is your age?

\noindent Q. (Optional) What gender do you identify with? Select all that apply.
\begin{itemize}
    \item Woman
    \item Man
    \item Non-binary
    \item Prefer not to disclose
    \item Prefer to self-describe:
\end{itemize}

\noindent Q. (Optional) What racial or ethnic groups best describe you? Select all that apply.
\begin{itemize}
    \item American Indian or Alaskan Native
    \item Asian
    \item Black or African American
    \item Hispanic and Latino
    \item Middle Eastern
    \item Native Hawaiian or Pacific Islander
    \item White/Caucasian
    \item Prefer not to disclose
    \item Prefer to self-describe:
\end{itemize}

\section{Interview Guide}
\label{sec:int_guide}
In this section, we include the general interview guide for the follow-up interviews.
Since questions could be based on the survey response, the questions varied slightly.
We displayed the participant's survey response on the screen while conducting the interview to help with recollecting the questions and their responses.

\noindent\underline{Context Gathering}
\begin{itemize}
    \item You mentioned in the survey that some disciplines that your work has involved are [disciplines reported in survey]. Which disciplines are you less familiar with, and why are you interested in these ``non-primary'' disciplines?
    \item What tasks does your work typically involve?
    \item What resources do you use for those tasks (non-AI and AI)?
\end{itemize}

\noindent\underline{Interdisciplinary Experiences and AI}
\begin{itemize}
    \item (When participant marked interdisciplinary experiences as challenging) Could you talk a bit about what is challenging about these experiences?
    \item (When participant marked interdisciplinary experiences as not challenging) How do you navigate these experiences?
    \item (Depending on participant's indicated use of AI) What type of AI information do you use?
    \item (Depending on participant's indicated use of AI) Why do you use AI and when you find it useful or beneficial?
    \item Is there anything that you use instead of AI or to complement/supplement information from AI?
    \item Are there times that you find AI unnecessary or challenging to use?
    \item How do you feel about the amount of control that you have over AI when you’re exploring unfamiliar information?
    \item Are there any gaps that AI doesn’t cover or creates when you’re working in an unfamiliar discipline?
\end{itemize}

\section{Non-digital Resources}
\begin{figure}[h]
    \centering
    \includegraphics[width=0.5\linewidth]{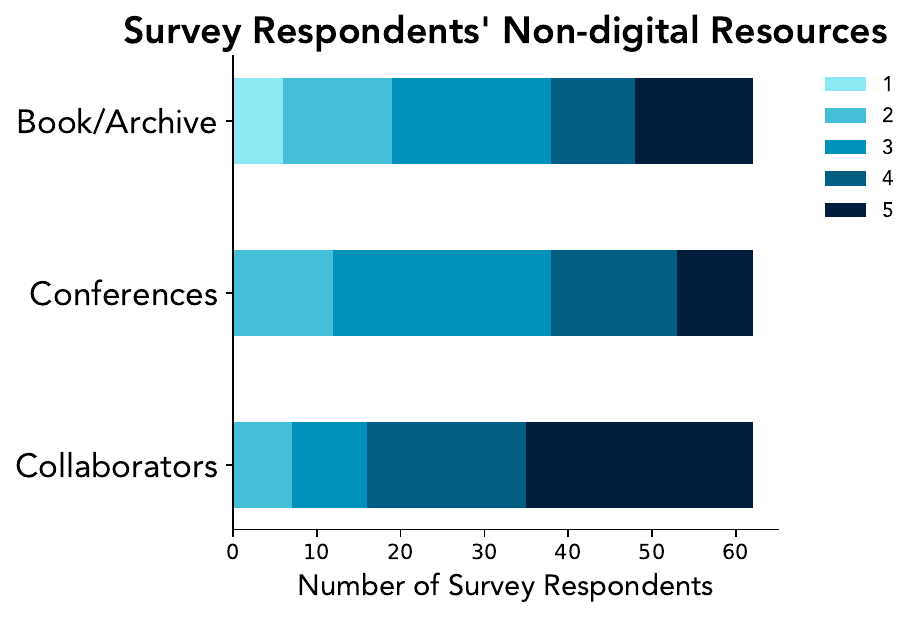}
    \caption{\textbf{An overview of respondents' use of non-digital resources.} The scale represents frequency of usage: 1 = Never, 2 = Rarely, 3 = Occasionally, 4 = Regularly, and 5 = Frequently.}
    \label{fig:nondigital}
    \Description{A plot with 3 bars showing the distributions of respondents' frequency of use of books/archives, conferences, and collaborators.}
\end{figure}

Figure \ref{fig:nondigital} shows the distributions of how often respondents use different non-digital resources; the corresponding question is in Appendix \ref{sec:survey_questions}.

\end{document}